\documentclass[fleqn]{article}
\usepackage{longtable}
\usepackage{graphicx}

\begin {document}
\begin{flushleft}
{\LARGE
{\bf Energy levels and radiative rates  for transitions in Ti X}
}\\

\vspace{1.5 cm}

{\bf {Kanti  M  ~Aggarwal and Francis  P   ~Keenan}}\\ 

\vspace*{1.0cm}

Astrophysics Research Centre, School of Mathematics and Physics, Queen's University Belfast, Belfast BT7 1NN, Northern Ireland, UK\\ 
\vspace*{0.5 cm} 

e-mail: K.Aggarwal@qub.ac.uk \\

\vspace*{1.50cm}

Received  13 May 2013\\
Accepted for publication 9 July 2013 \\
Published xx  Month 2013 \\
Online at stacks.iop.org/PhysScr/vol/number \\

\vspace*{1.5cm}

PACS Ref: 31.25 Jf, 32.70 Cs,  95.30 Ky

\vspace*{1.0 cm}

\hrule

\vspace{0.5 cm}
{\Large {\bf S}} This article has associated online supplementary data files \\
Tables 3 and 4 are available only in the electronic version at stacks.iop.org/PhysScr/vol/number/mmedia

\end{flushleft}

\clearpage


\begin{abstract}

We report calculations of energy levels, radiative rates, oscillator strengths and line strengths for transitions among the lowest 345 levels of Ti X. These include 146 levels of the $n \le$ 3 configurations and  86 of 3s$^2$4$\ell$,  3s$^2$5$\ell$ and 3s3p4$\ell$,   plus some of the 3s$^2$6$\ell$,  3p$^2$4$\ell$ and 3s3p5$\ell$ levels. The general-purpose relativistic atomic structure package ({\sc grasp}) and flexible atomic code ({\sc fac}) are adopted for the calculations. Radiative rates, oscillator strengths and line strengths are provided for all electric dipole (E1), magnetic dipole (M1), electric quadrupole (E2) and magnetic quadrupole (M2) transitions among the 345 levels, although calculations have been performed for a much larger number of levels.  Comparisons are made with existing results and the accuracy of the data is assessed. Additionally, lifetimes for all 345 levels are listed. Extensive comparisons of lifetimes are made for the lowest 40 levels, for which discrepancies with recent theoretical work are up to 30\%. Discrepancies in lifetimes are even larger, up to a factor of four, for higher excited levels. Furthermore, the effect of large CI is found to be insignificant for both the energies and  lifetimes for the lowest 40 levels of Ti X which belong to the 3s$^2$3p, 3s3p$^2$, 3s$^2$3d,  3p$^3$ and 3s3p3d configurations. However, the contribution of CI is more appreciable for the energy levels and radiative rates among higher excited levels. Our listed energy levels are estimated to be accurate to better than 1\% (within 0.1 Ryd), whereas results for other parameters are probably accurate to  better than 20\%. 
 
\end{abstract}

\clearpage

\section{Introduction}

Iron group elements (Sc - Zn) are becoming increasingly important in the study of astrophysical plasmas, as many of their  lines are frequently observed from different ionisation stages. These lines provide a wealth of data about the plasma characteristics, including  temperature, density and chemical composition. More importantly, iron group elements are often impurities in fusion reactors, and to estimate the power loss from the impurities, atomic data (including energy levels and  oscillator strengths or radiative decay rates) are required for many ions. The need for atomic data has become even greater with the developing  ITER project. Since there is a paucity of measured parameters, one must depend on theoretical results. Therefore,  recently we have reported atomic parameters for many ions of the iron group elements -- see for example  \cite{fe26}--\cite{adt1}  and references therein. Among Ti ions, results have already been provided for Ti XXII \cite{ti22}, Ti XXI \cite{ti21}, Ti XX \cite{ti20}, Ti XIX \cite{ti19} and Ti VI \cite{ti6}, and here we focus our attention on Al-like Ti X.

Several emission lines of Ti ions have been observed in astrophysical plasmas, as listed in the CHIANTI database at ${\tt {\verb+http://www.chiantidatabase.org+}}$. For Ti X, Edlen \cite{be} was the first to classify two lines  followed by \cite{gfj} and \cite{fp}, who identified several mutiplets among the low-lying terms. However, we are not aware of any astrophysical observations for Ti X, although many emission lines are listed in the 70-35,700 ${\rm \AA}$ wavelength range in the {\em Atomic Line List} (v2.04) of Peter van Hoof at ${\tt {\verb+http://www.pa.uky.edu/~peter/atomic/+}}$, because these are useful in the generation of synthetic spectra. Furthermore, laboratory measurements for  lines of Ti X were made as early as 1969 \cite{se}, which have been analysed by Ekberg and Svensson \cite{es}. Further measurements were made by Smitt {\em et al} \cite{sso} and Churilov and Levashov \cite{cl}. The experimental data have been  compiled by  Corliss and Sugar \cite{cs} and are also available at the NIST (National Institute of Standards and Technology)  website {\tt http://www.nist.gov/pml/data/asd.cfm}.

Considering the importance of Ti ions, several calculations have been performed for  Ti X -- see for example \cite{bcf} -- \cite{mm1}. A variety of sophisticated methods and codes have been adopted by these workers, but the calculations of \cite{bcf}, \cite{knh}, \cite{cff1} and \cite{sit} have  been confined to the lowest 40 levels of Ti X belonging to the 3s$^2$3p, 3s3p$^2$, 3s$^2$3d,  3p$^3$ and 3s3p3d configurations. However, in  plasma modelling atomic data for a larger number of levels are required, because of the cascading effect. Therefore, Gupta and Msezane \cite{gm} performed a larger calculation for 110 levels, the additional 70  arising from the 3p$^2$3d, 3s3d$^2$, 3p3d$^2$, 3s$^2$4$\ell$  and 3s3p4s configurations. However, these 8 configurations generate a total of 103 levels, i.e. Gupta and Msezane excluded 33 levels. For example, they identified only half of the 3p$^2$3d levels, as may be seen in Table 4 of \cite{mm1} or the current Table 2a. The latest calculations in the literature are by Singh {\em et al} \cite{mm1}, who not only included extensive CI (configuration interaction), but also reported energies and radiative rates for transitions among 303 levels, which belong to the (1s$^2$2s$^2$2p$^6$) 3s$^2$3p, 3s3p$^2$, 3s$^2$3d, 3p$^3$, 3s3p3d, 3p$^2$3d, 3s3d$^2$,
3p3d$^2$, 3d$^3$, 3s$^2$4$\ell$, 3p$^2$4$\ell$, 3s3p4$\ell$, 3s$^2$5$\ell$, and 3s$^2$6$\ell$ configurations. We discuss these below.

Singh {\em et al} \cite{mm1} adopted the CIV3 code of Hibbert \cite{civ3} and included one-body relativistic operators in their calculations, which should be sufficient for a moderately heavy ion such as Ti X. Furthermore, they included extensive CI with 80 configurations ($n \le$ 6, $\ell \le$ 3 -- see their Table 3)  to accurately determine the energy levels. In addition, they {\em adjusted} the Hamiltonian in accordance with the NIST compilations (a process known as ``fine-tuning"), which minimises the differences between theoretical and experimental energy levels.   Therefore, their reported energy levels should be the best currently  available. However, there are some problems with their results. Firstly, the process of fine-tuning is only helpful if experimental data are available for a majority of levels, which is {\em not} the case with Ti X as may be noted from Table 4 of \cite{mm1}. For the same reason it does not correct the level orderings. Secondly, if energy levels are accurate then the subsequent results for radiative rates (A- values) are likely to be equally accurate, which is not  the case in their calculations, because the differences between their results and those compiled by NIST are up to an order of magnitude for several transitions, as can be seen in their Table C. Subsequently, a closer examination of their lifetimes \cite{mm2}  reveals significant differences not only with the measurements but also with other theoretical results -- see their Table 2 or the present Tables 7 and 8. Therefore, there is definitely scope for improvement over their results, particularly for the A- values. Thirdly, they have reported energies for 303 levels, i.e. up to  3s$^2$6f, and our calculations show that to cover the highest energy of these levels, we need to span over 339 levels. The missing 36 levels particularly affect the calculations of lifetimes. Finally, and most importantly, they have reported A- values only for electric dipole (E1) transitions, whereas in plasma modelling the A- values are also required for the electric quadrupole (E2), magnetic dipole (M1), and magnetic quadrupole (M2) transitions,  as demonstrated by Del Zanna {\em et al}\,  \cite{del04}.  These transitions also affect the subsequent calculations of lifetimes, particularly of those which do not connect with the E1 transitions. For this reason, in a subsequent paper Singh {\em et al} \cite{mm2} reported lifetimes for only 294 levels among 303. Therefore, our {\em aim} is to improve upon the calculations of Singh {\em et al} \cite{mm1}, so that the atomic data can be confidently applied to the modelling of plasmas.

For our calculations we have adopted the {\sc grasp} (general-purpose relativistic atomic structure package) code to generate the wavefunctions. This code was originally developed as GRASP0 by Grant {\em et al} \cite{grasp0} and has been updated by Dr. P. H. Norrington ({\tt http://web.am.qub.ac.uk/DARC/}). It is fully relativistic, and is based on the $jj$ coupling scheme. Further relativistic corrections arising from the Breit interaction and QED (quantum electrodynamics) effects have also been included. Additionally, we have used the option of {\em extended average level} (EAL),  in which a weighted (proportional to 2$j$+1) trace of the Hamiltonian matrix is minimised. This produces a compromise set of orbitals describing closely-lying states with moderate accuracy, and generally yields results comparable to other options, such as {\em average level} (AL), as noted by Aggarwal {\em et al}  for several ions of Kr \cite{kr} and Xe \cite{xe}.  

\section{Energy levels}

Although  Ti X is moderately heavy ($Z$ = 22) and 9 times ionized, CI is still very important for an accurate determination of energy levels. For this reason most earlier workers have included CI with additional configurations. Following some tests with a number of $n \le$ 6 configurations, we  have also arrived at the same conclusion that an elaborate CI needs to be included to achieve a better accuracy in the determination of energy levels. Therefore, we have performed a series of calculations  with increasing amount of CI, and have determined  that the most important configurations which need to be included in a calculation are: 3s$^2$3p, 3s3p$^2$, 3s$^2$3d, 3p$^3$, 3s3p3d, 3p$^2$3d, 3s3d$^2$, 3p3d$^2$, 3d$^3$, 3s$^2$4$\ell$, 3s3p4$\ell$, 3s3d4$\ell$, 3s3p5$\ell$, 3s$^2$5$\ell$, 3s$^2$6$\ell$ ($\ell \le$ f) and 3p$^2$4$\ell$. These 39 configurations generate 530 levels in total and closely interact and intermix. The highest energy range for these levels is up to $\sim$ 15 Ryd, and we refer to this calculation as GRASP1. However,  to assess the impact of additional CI, we have performed another calculation (GRASP2) including the 3p3d4$\ell$, 3s3d5$\ell$ and 3p3d5$\ell$ configurations. These additional 14 configurations generate a  further 857 levels (1387 in total) and their energies are {\em above} 13.5 Ryd, i.e. there is no strong interaction with the lowest $\sim$ 300 levels which are of interest here. Before we discuss our results, we note that calculations have also been performed with  configurations such as 3p$^2$5$\ell$ and 3p$^2$6$\ell$, included by Singh {\em et al} \cite{mm1}, but are not discussed here because they generate levels with much higher energies, and their impact on the levels considered  is insignificant. Nevertheless, their impact will be discussed later.

\subsection{Lowest 40 levels}

In Table 1 we list our calculated energies with the  {\sc grasp} code for the lowest 40 levels of the 3s$^2$3p, 3s3p$^2$, 3s$^2$3d,  3p$^3$ and 3s3p3d configurations. Results from both calculations with the {\sc grasp} code (GRASP1 and GRASP2) described above  are listed here, and include  the Breit and QED corrections, which have shifted the ground level energy by 0.85 Ryd. For excited levels,  the Breit contribution is up to 0.01 Ryd, depending on the level,  whereas that of QED is $\le$ 0.002 Ryd. Also included in this table are the experimental energies compiled by NIST and the theoretical values obtained by  Froese Fischer {\em et al} \cite{cff1}, Santana {\em et al} \cite{sit} and Singh {\em et al} \cite{mm1} from the MCHF (multi configuration Hartree-Fock), MPPT (M{\o}ller-Plesset perturbation theory) and  CIV3  (configuration interaction version 3) codes, respectively. 

Theoretical energies obtained by Singh {\em et al} \cite{mm1} are in close agreement with those of NIST, because of the adjustments made. The MCHF energies \cite{cff1} are also in close agreement with the NIST compilations, but their energy for 3s$^2$3p $^2$P$^o_{3/2}$ (level 2) is {\em lower} by $\sim$ 3\%. Their level orderings are also nearly the same as of NIST. Similarly, the MPPT energies \cite{sit} agree closely with those of NIST, but the level orderings are different in a few instances, particularly for the [3s3p($^1$P)3d] $^2$P$^o_{1/2,3/2}$ and  $^2$D$^o_{3/2,5/2}$ levels. We also note that in Table 4 of \cite{sit} the labelings  are incorrect for the last 10 levels. Our energies obtained with the {\sc grasp} code agree closely with the NIST compilations, but only for the lowest $\sim$ 30 levels. For higher levels our energies are higher than those of NIST by up to 0.1 Ryd (2\%), and the effect of the additional CI included in the GRASP2 calculations is negligible for these lowest 40 levels, in both magnitude as well as the orderings. Finally, all three calculations with the MCHF, MPPT and {\sc grasp} codes give the same orderings for the [3s3p($^3$P)3d] $^4$D$^o_{1/2}$ and  $^4$P$^o_{1/2}$ levels (24 and 25), and differ with the NIST compilation, although the energy difference between these two levels is very small ($\sim$ 0.02 Ryd).  

To further assess the accuracy of our results, we have performed another calculation with the  {\em Flexible Atomic Code} ({\sc fac}) of Gu \cite{fac},  available from the website {\tt http://kipac-tree.stanford.edu/fac}. This is also a fully relativistic code which provides a variety of atomic parameters, and yields results for energy levels and A- values comparable to {\sc grasp}, as already shown for several other ions, see for example:  Aggarwal {\em et al} \cite{fe15} for  Mg-like ions and \cite{ti22}--\cite{ti6} for Ti ions. In addition, a clear advantage of this code is its high efficiency which means that large calculations can be performed within a reasonable time frame of a few weeks. Thus results from {\sc fac} will be helpful in assessing the accuracy of our energy levels (particularly the higher ones to be discussed later) and radiative rates.

As with the {\sc grasp} code, we have performed a series of calculations using {\sc fac} with increasing amounts of CI. However, here we focus on only two calculations, namely (i) FAC1, which includes the 1387 levels of the GRASP2 calculations, plus 4 levels of the (3s$^2$) 6g and 6h configurations, and (ii) FAC2, which includes a total of 12,139 levels, which arise from all possible combinations of the $n$ = 3 configurations {\em and} (3*2) 4*1, 5*1, 6*1, (3*1) 4*2, 5*2 and 6*2 configurations. The results obtained from these two calculations are also listed in Table 1. Both calculations yield the same {\em orderings} as in our present results with {\sc grasp},  and agree with each other within 0.01 Ryd. This indicates that the inclusion of larger CI in FAC2 is of no significance. The agreement between the {\sc grasp} and {\sc fac} calculations is also within 0.01 Ryd, which is highly satisfactory. Thus all calculations give comparable energies in magnitude. The orderings of the lowest 40 levels of Ti X are the same in our two calculations with the {\sc grasp} and {\sc fac} codes, but all other theoretical or experimentally-compiled orderings slightly differ with one another.

\subsection{$n$ = 3 levels}

Apart from the lowest 40 levels listed in Table 1, the other possible configurations within the $n$ = 3 complex are: 3p$^2$3d, 3s3d$^2$, 3p3d$^2$ and 3d$^3$, which generate 108 levels in total. In Table 2a we compare our energies, for 28 levels of the 3p$^2$3d configuration, from the {\sc grasp} and {\sc fac} codes with the NIST compilations, and the only other available results of Gupta and Msezane \cite{gm} and Singh {\em et al} \cite{mm1}, who have both adopted the CIV3 code \cite{civ3}. The NIST energies are not available for all the levels and thus necessitate  the use of theoretical results in plasma modelling. Our energies from the GRASP1 calculations  have the same ordering as NIST, but differ in magnitude by up to 0.15 Ryd, particularly for the higher levels. Inclusion of larger CI in the GRASP2 calculation, lowers the energies by up to 0.04 Ryd, and the  ordering of the 3p$^2$($^1$D)3d $^2$S$_{1/2}$ and 3p$^2$($^1$S)3d $^2$D$_{3/2}$ levels is reversed. However, differences of  $\sim$ 0.1 Ryd remain with the NIST compilations. Our FAC1 energies are also comparable with the GRASP2 calculations (within 0.01 Ryd), and inclusion of larger CI in FAC2 does not affect  levels of the 3p$^2$3d configuration. Therefore, we have confidence in our results.

Gupta and Msezane \cite{gm} identified only half of the levels of the 3p$^2$3d configuration, and among these misidentified two, namely 3p$^2$($^1$D)3d $^2$P$_{1/2}$ and 3p$^2$($^3$P)3d $^2$P$_{1/2}$ (i.e. levels 11 and 20). Furthermore, in spite of adjusting their Hamiltonian in accordance with the experimental energies, their results differ with the NIST compilations by up to 0.1 Ryd for several levels, and hence are neither complete nor very accurate.  Similarly,  the energies reported by Singh {\em et al} \cite{mm1} differ from the NIST values by up to 0.1 Ryd for a few levels, see for example the last two, i.e. 3p$^2$($^3$P)3d $^2$D$_{5/2,3/2}$. Furthermore, their ordering is slightly different for a few levels, such as 7, 12 and 23. However, before we draw any firm conclusion we compare energies for the levels of some other configurations. 

In Table 2b we compare our energies for levels of the 3s3d$^2$ configuration from our calculations with {\sc grasp} (GRASP1 and GRASP2) and {\sc fac} (FAC1 and FAC2). Also included in this table are  results from the NIST compilations, and the earlier calculations of Gupta and Msezane \cite{gm} and Singh {\em et al} \cite{mm1} with the CIV3 code. NIST energies are available for only 3 of the 16 levels, and those of Singh {\em et al} show no discrepancy because of the adjustments made. For most of the levels (except $^2$S$_{1/2}$) the two CIV3 calculations are comparable, and are closer to our energies from the GRASP1 calculations. However, a larger CI included in the GRASP2 calculations lowers the energies by up to 0.1 Ryd. Our GRASP2 and FAC1 energies are comparable (within 0.02 Ryd), but an extensive CI in FAC2 lowers the energies further by 0.02 Ryd. Thus the differences between the GRASP2 and FAC2 calculations are up to 0.04 Ryd, and indicate the growing importance of CI.

In Table 2c we compare our energies for levels of the 3p3d$^2$ configuration from {\sc grasp} (GRASP1 and GRASP2)  and {\sc fac}  (FAC1 and FAC2) with those of Gupta and Msezane \cite{gm} and Singh {\em et al} \cite{mm1} from the CIV3 code. NIST have not compiled energies for these levels, and both calculations with CIV3 are incomplete as they do not cover all levels of the 3p3d$^2$ configuration. Differences between the two CIV3 calculations \cite{gm},\cite{mm1} are up to 0.15 Ryd for several levels, and their orderings also differ in a few instances. In particular, Gupta and Msezane  interchanged the ordering of the 3p3d$^2$($^3$F) $^2$F$^o_{5/2,7/2}$ and 3p3d$^2$($^1$D) $^2$F$^o_{5/2,7/2}$ levels, although these (and many other) levels are highly mixed -- see Table II of \cite{gm} and/or Table 4 of \cite{mm1}.

There is no discrepancy between our GRASP1 and GRASP2 energies, as both agree to within 0.04 Ryd and have the same ordering. This indicates that the effect of additional CI included in the GRASP2 calculations is of no particular significance as far as the levels of the 3p3d$^2$ configuration are concerned. However, a much larger CI included in the FAC2 calculations does lower the energies, by a maximum of 0.1 Ryd, particularly for a few higher levels (38--45). Similarly, the CIV3 energies of  Gupta and Msezane \cite{gm} are comparable to our GRASP1 calculations for most of the levels, but it is surprising to see the discrepancy with the other CIV3 results of Singh {\em et al} \cite{mm1}, which are the {\em highest} among all data listed in Table 2c. This is in spite of the fact that they have included a large CI, but for these levels there are no energies by NIST with which they could make an adjustment. Therefore, a comparative inaccuracy in their energy levels is becoming apparent. 

Finally, in Table 2d we compare our energies from the {\sc grasp} and {\sc fac} codes with the only other available results of Singh {\em et al} \cite{mm1} for the levels of the 3d$^3$ configuration. Our GRASP1 and GRASP2 energies differ by up to 0.1 Ryd for several levels and the orderings are also different in a few instances -- see for example, the 3d$^3$($^4$P) $^4$P$_{1/2,3/2,5/2}$ levels. Clearly, the additional CI included in the GRASP2 calculations has improved upon the results obtained in GRASP1. However, further CI included in the FAC2 calculations is of no significant advantage, as results with FAC1 agree within 0.03 Ryd for all levels. On the other hand, the CIV3 energies of Singh {\em et al} are higher by up to 0.3 Ryd for several of the levels, and are therefore not as accurate as expected.

To conclude, we may say that our energies with the GRASP2 and FAC1 calculations are comparable for a majority of the $n$ = 3 levels, but the additional CI included in the FAC2 calculations has improved upon the energies as well as the ordering of (some of) the levels. Finally, in comparison to a variety of calculations, the energy levels of Singh {\em et al} \cite{mm1} from the CIV3 code are neither complete nor very accurate.

\subsection{3s3p4$\ell$ levels}

In Table 2e we compare energies for the levels of the 3s3p4$\ell$ configurations of Ti X. Included in this table are energies from our calculations with  {\sc grasp} (GRASP1 and GRASP2) and {\sc fac} (FAC1 and FAC2),  plus the earlier CIV3 results of Gupta and Msezane \cite{gm} and  Singh {\em et al} \cite{mm1}. Energies for the 3s3p($^3$P)4s $^4$P$^o_{1/2,3/2,5/2}$ levels  are also available on the NIST website, which are closer to our GRASP1 results (within $\sim$ 0.01 Ryd). Our GRASP2 energies, obtained with a larger CI, are lower than those from GRASP1 by up to 0.15 Ryd for several levels, such as 21--25 and 43--48, and are particularly lower for the 3s3p($^1$P)4f $^2$G$_{7/2,9/2}$ levels (by 0.5 Ryd). Thus the effect of larger CI is more pronounced on these two levels than other listed in Table 2e. Furthermore, there is no discrepancy between the GRASP2 and FAC1 calculations, because both include the same CI. However, a larger calculation performed in FAC2 lowers the energies further by 0.05 Ryd, particularly for the 3s3p($^1$P)4f $^2$G$_{7/2,9/2}$ levels. For other levels the effect of larger CI is much smaller.

The CIV3 energies of Gupta and Msezane \cite{gm} are available  only for the 7 levels of the 3s3p4s configuration for which there are no major discrepancies with our calculations. However, the corresponding CIV3 energies of Singh {\em et al} \cite{mm1}, available for a majority of the levels listed in Table 2e, differ from our GRASP1 calculations, by up to 0.4 Ryd for several levels (see for example levels 57--72). Since they have also included a large CI in their calculations, it will be fairer to compare with our other results, i.e. GRASP2, FAC1 and FAC2. However, the differences with these calculations are also up to 0.25 Ryd for several levels. In some cases the energies of  Singh {\em et al} are lower (such as levels 18, 19, 36 and 57--70), but in others are higher (such as 6, 7, 37--39 and 45--48). Hence  there is no consistency in the behaviour of differences, but in general their results obtained for the 3s3p($^3$P)4$\ell$ levels are comparatively more accurate than for the 3s3p($^1$P)4$\ell$ levels.

\subsection{Lowest 345 levels}

In Table 3 (see supplementary data, available online at stacks.iop.org/PhysScr/vol/number/mmedia) we list our final energies, in increasing order,  obtained using the {\sc grasp} code with CI among 53 configurations listed in  section 2, which correspond to the GRASP2 calculations. These configurations generate 1387 levels, but for conciseness energies are listed only for the lowest 345 levels, which include 146 levels of the $n \le$ 3 configurations and  86 of 3s$^2$4$\ell$,  3s$^2$5$\ell$ and 3s3p4$\ell$,   plus some of the 3s$^2$6$\ell$,  3p$^2$4$\ell$ and 3s3p5$\ell$ levels.  However, energies corresponding to any of the calculations described in section 2.1 and for any desired number of levels up to 12,139 can be obtained on request from the first author (K.Aggarwal@qub.ac.uk).

Although calculations with the {\sc fac} code have been performed with the inclusion of larger CI, energies obtained with the {\sc grasp} code alone are listed in Table 3 (see supplementary data, available online at stacks.iop.org/PhysScr/vol/number/mmedia). This is partly because both codes provide energies with comparable accuracy as demonstrated and discussed in sections 2.1 to 2.3, but mainly because the $LSJ$ designations of the levels are also determined  in the {\sc grasp} code. For a majority of users these designations are more familiar and hence preferable. However, we note that the $LSJ$ designations provided in this table are not always unique, because some of the levels are highly mixed, mostly from the same but sometimes with other configurations. This has also been discussed by  Gupta and Msezane \cite{gm} and  Singh {\em et al} \cite{mm1}. Therefore, care has been taken to provide the most appropriate designation of a level/configuration, but a redesignation of  these cannot be ruled out in a few cases. 

For the 345 levels listed in Table 3 (see supplementary data, available online at stacks.iop.org/PhysScr/vol/number/mmedia), comparisons with the NIST compilations of experimental energies has been possible for only a few. There are no major discrepancies  with our calculations, although the orderings of the levels differ in a few instances. However, extensive comparisons have been possible, for a majority of the levels, with  other available theoretical work, particularly of  Singh {\em et al} \cite{mm1},  as  shown in Tables 1 and 2 (a--e). Based on these comparisons it is concluded that CI is very important for the energy levels of Ti X, but mostly among those configurations whose levels interact closely.  Singh {\em et al} also included a large CI in their calculations with the CIV3 code \cite{civ3}, but some of the configurations they considered, such as 3p$^2$5$\ell$ and 3p$^2$6$\ell$, are not of  importance as noted in section 1 and discussed in section 2. Similarly, they adjusted their calculated energies using the NIST compilations, but this has not been useful as  experimental energies are not available for a majority of the levels. For these reasons, differences between our calculations and those reported by Singh {\em et al} are significant (up to 0.5 Ryd) for many levels, and level orderings also differ in a few instances. On the other hand, our GRASP2, FAC1 and FAC2 energies are comparable for most of the levels, in both magnitude as well as orderings. Thus we have confidence in our results, and based on a variety of comparisons assess the accuracy of our energy levels listed in Table 3 (see supplementary data, available online at stacks.iop.org/PhysScr/vol/number/mmedia)  to be better than 1\%. 

\section{Radiative rates}

The absorption oscillator strength ($f_{ij}$) and radiative rate A$_{ji}$ (in s$^{-1}$) for a transition $i \to j$ are related by the following expression:

\begin{equation}
f_{ij} = \frac{mc}{8{\pi}^2{e^2}}{\lambda_{ji}}^2 \frac{{\omega}_j}{{\omega}_i}A_{ji}
 = 1.49 \times 10^{-16} \lambda^2_{ji} (\omega_j/\omega_i) A_{ji}
\end{equation}
where $m$ and $e$ are the electron mass and charge, respectively, $c$ is the velocity of light, 
$\lambda_{ji}$ is the transition energy/wavelength in $\rm \AA$, and $\omega_i$ and $\omega_j$ are the statistical weights of the lower $i$ and upper $j$ levels, respectively.
Similarly, the oscillator strength $f_{ij}$ (dimensionless) and the line strength $S$ (in atomic unit, 1 a.u. = 6.460$\times$10$^{-36}$ cm$^2$ esu$^2$) are related by the 
following standard equations:

\begin{flushleft}
for the electric dipole (E1) transitions: 
\end{flushleft} 
\begin{equation}
A_{ji} = \frac{2.0261\times{10^{18}}}{{{\omega}_j}\lambda^3_{ji}} S \hspace*{1.0 cm} {\rm and} \hspace*{1.0 cm} 
f_{ij} = \frac{303.75}{\lambda_{ji}\omega_i} S, \\
\end{equation}
\begin{flushleft}
for the magnetic dipole (M1) transitions:  
\end{flushleft}
\begin{equation}
A_{ji} = \frac{2.6974\times{10^{13}}}{{{\omega}_j}\lambda^3_{ji}} S \hspace*{1.0 cm} {\rm and} \hspace*{1.0 cm}
f_{ij} = \frac{4.044\times{10^{-3}}}{\lambda_{ji}\omega_i} S, \\
\end{equation}
\begin{flushleft}
for the electric quadrupole (E2) transitions: 
\end{flushleft}
\begin{equation}
A_{ji} = \frac{1.1199\times{10^{18}}}{{{\omega}_j}\lambda^5_{ji}} S \hspace*{1.0 cm} {\rm and} \hspace*{1.0 cm}
f_{ij} = \frac{167.89}{\lambda^3_{ji}\omega_i} S, 
\end{equation}

\begin{flushleft}
and for the magnetic quadrupole (M2) transitions: 
\end{flushleft}
\begin{equation}
A_{ji} = \frac{1.4910\times{10^{13}}}{{{\omega}_j}\lambda^5_{ji}} S \hspace*{1.0 cm} {\rm and} \hspace*{1.0 cm}
f_{ij} = \frac{2.236\times{10^{-3}}}{\lambda^3_{ji}\omega_i} S. \\
\end{equation}

The A- and f- values have been calculated in both Babushkin and Coulomb gauges, which are  equivalent to the length and velocity forms in the non-relativistic nomenclature. However, the results are presented here in the length form alone  which are considered to be comparatively more accurate  \cite{ipg} -- \cite{jps}.   In Table 4 (see supplementary data, available online at stacks.iop.org/PhysScr/vol/number/mmedia) we present transition energies ($\Delta$E$_{ij}$ in ${\rm \AA}$), radiative rates (A$_{ji}$ in s$^{-1}$), oscillator strengths ($f_{ij}$, dimensionless), and line strengths ($S$ in a.u.) for all 18,267 electric dipole (E1) transitions among the lowest 345 levels of Ti X. The {\em indices} used to represent the lower and upper levels of a transition have already been defined in Table 3 (see supplementary data, available online at stacks.iop.org/PhysScr/vol/number/mmedia). Also, in calculating the  above parameters we have used the Breit and QED-corrected theoretical energies/wavelengths as listed in Table 3. However, only A- values are included in Table 4 for the 24,034 electric quadrupole (E2), 18,131  magnetic dipole (M1), and 24,098 magnetic quadrupole (M2) transitions. Corresponding results for f- or S- values can be easily obtained by using Eqs. (1-5). 

In Table 5  we compare our A- values for transitions among the lowest 40 levels  from the calculations with {\sc grasp} (GRASP1 and GRASP2) and  {\sc fac} (FAC1 and FAC2), with those of Froese-Fischer {\em et al} \cite{cff1} and Singh {\em et al} \cite{mm1} from the  MCHF and CIV3 codes, respectively. Also included in the table are A- values compiled by NIST and f- values from our GRASP2 calculations, as they provide an indication of the strength of a transition. For almost all transitions,  there is close agreement among A- values from the GRASP1, GRASP2, FAC1 and FAC2 calculations, with the only exceptions being 4--24 and 12--33, for which the A- values differ by up to a factor of two. However,  these are weak transitions with f = 0.0073 and 0.0003, respectively. Similarly, there are no discrepancies with the A- values of NIST. Agreement with the MCHF A- values is also within $\sim$ 10\% for all transitions, except one, namely 4--24 (3s3p$^2$ $^4$P$_{3/2}$ -- 3s3p($^3$P)3d $^4$D$^o_{1/2}$). This is a weak transition with f = 0.0073 and our A- values from GRASP1, GRASP2 and FAC1 differ with up to 50\%. However, the MCHF A -value for this transition is in excellent agreement with our FAC2 calculation and indicates the importance of larger CI for some of the transitions.

On the other hand, the CIV3 A- values of Singh {\em et al} differ by up to a factor of two for several transitions, such as: 1--8/9, 2--8/9 and 7--13. Moreover, discrepancies between their A- values and our calculations and NIST compilations are up to an order of magnitude for a few transitions, such as: 7--36 and 11--31. These large differences are in spite of the fact that their energies for the lowest 40 levels of Ti X are comparatively in better agreement with the NIST compilations, as shown in Table 1. A normal practice in a CIV3 calculation is to first survey all levels of a configuration and then eliminate those whose eigenvectors are below a certain magnitude (say $\sim$ 0.2) before performing a final run for transition rates. This exercise is undertaken to keep the calculations manageable within the limited computational resources available, and is the most likely reason for the differences in A- values between our elaborate calculations with the {\sc grasp} and {\sc fac} codes and those of Singh {\em et al} with CIV3. Similar differences, and for the same reasons, were noted by Aggarwal {\em et al} \cite{feix} in their calculations for transitions in Fe IX  \cite{mm3}. Therefore, as noted earlier \cite{fst}, we emphasise once again that the process of fine-tuning may make the theoretical energy levels more accurate in magnitude, but {\em not} the subsequent calculations of A- values (or other parameters such as lifetimes and collision strengths), if inherent deficiencies are already present. Based on the comparisons shown in Table 5 for several transitions, strong as well as weak, we can confidently state that the A- values reported by Singh {\em et al} \cite{mm1} are not as accurate as expected, and differ for several transitions by up to an order of magnitude.

In Table 6 we compare our A- values from calculations with {\sc grasp} (GRASP1 and GRASP2) and {\sc fac} (FAC1 and FAC2) for transitions from  the ground configuration (3s$^2$3p  $^2$P$^o_{1/2,2/2}$) to higher excited levels of Ti X. Our GRASP1 calculations include minimum CI, and as a result the discrepancies for the A-  values with other calculations are up to three orders of magnitude, particularly for  comparatively weaker transitions, such as: 1--60/61/62/63 and 2--61/62/63/64. This confirms, once again,  the importance of CI in the determination of A- values, as for the energy levels. However, agreement among the other three calculations, namely GRASP2, FAC1 and FAC2, is within $\sim$ 20\% for most of the transitions, although for some weaker transitions (such as: 1--60/85 and 2--60) the discrepancies are larger (up to a factor of three). Generally,  f- values for weaker transitions are less accurate, because mixing coefficients from several components may have an additive or cancellation effect, which affects the weaker transitions more than the strong ones. Overall, we may state that the  A- values obtained from our GRASP2 and/or FAC1 calculations are as accurate as those from the FAC2 calculations, for a majority of (strong) transitions.

One of the general criteria to assess the accuracy of radiative rates is to compare the length and velocity forms of the f- or A- values. However, such comparisons are only desirable, and are {\em not} a fully sufficient test to assess accuracy, as calculations based on different methods (or combinations of configurations) may give comparable f- values in the two forms, but entirely different results in magnitude. Generally, there is a good agreement between the length and velocity forms of the f- values for {\em strong} transitions (f $\ge$ 0.01), but differences  between the two can sometimes be substantial even for some very strong transitions, as demonstrated through various examples by Aggarwal {\em et al} \cite{fe15}. Nevertheless, for almost all of the strong E1 transitions  the two forms agree to within 20\%, but the differences for 321 ($<$2\%) of the transitions are slightly larger. In fact, for only 40 transitions do the f- values differ by over  50\% (mostly within a factor of three), and for four transitions (128--306, 132--307, 239--250 and 240--249) the two forms differ by  up to an order of magnitude. Therefore, on the basis of these and earlier comparisons shown in Tables 5 and 6 we may state that for a majority of the strong E1 transitions, our radiative rates are accurate to better than 20\%. However, for the weaker transitions this assessment of accuracy does not apply, because such transitions are very sensitive to mixing coefficients, and hence differing amount of CI (and methods) produce different A- values, as discussed in detail by Hibbert \cite{ah3}. This is the main reason that the two forms of f- values for some weak transitions differ significantly (by orders of magnitude), and examples include 1--227 (f  = 1.0$\times$10$^{-9}$), 4--122 (f = 2.7$\times$10$^{-6}$) and 5--178 (f = 1.2$\times$10$^{-7}$). The f- values for weak transitions may be required in plasma modelling for completeness, but their contributions are less important in comparison to stronger transitions with f $\ge$ 0.01. For this reason many authors (and some codes) do not normally report A- values for very weak transitions.
 
\section{Lifetimes}

The lifetime $\tau$ of a level $j$ is defined as follows:

\begin{equation}
{\tau}_j = \frac{1}{{\sum_{i}^{}} A_{ji}}.
\end{equation}

In Table 3 (see supplementary data, available online at stacks.iop.org/PhysScr/vol/number/mmedia)  we include lifetimes for all 345 levels from our calculations with the {\sc grasp} code (corresponding to GRASP2). These results {\em include} A- values from all types of transitions, i.e. E1, E2, M1 and M2. Most of the earlier theoretical or experimental results for $\tau$ are confined to the lowest 40 levels of Ti X, as mentioned in section 1. Therefore, we focus our efforts primarily on these levels, although Singh {\em et al} \cite{mm2} have reported theoretical lifetimes for 294 levels of Ti X.

In Table 7 we list  lifetimes from our calculations with  {\sc grasp} (GRASP1 and GRASP2) and {\sc fac} (FAC1 and FAC2) as described in section 2. Also included for comparison are the experimental lifetimes of Pinnington {\em et al} \cite{ehp1}, \cite{ehp2} and Tr{\" a}bert {\em et al} \cite{et1}, and the theoretical results of Singh {\em et al} \cite{mm2} from the CIV3 code, Froese-Fischer {\em et al} \cite{cff1} from the MCHF code, and  Safronova {\em et al} \cite{uis} from many-body perturbation theory (MBPT). Additionally, we have included the contributions from four types of transitions, i.e. electric dipole (E1), electric quadrupole (E2), magnetic dipole (M1), and magnetic quadrupole (M2) in all our calculations from the {\sc grasp} and {\sc fac} codes, although the E1 transitions alone are dominant (and hence sufficient) for the lowest 40 levels. Before we discuss the results for lifetimes, we note that the reported results of Pinnington {\em et al} \cite{ehp1} are uncertain, mainly due to the (mis)identification of transition assignments, as indicated in a later work by Pinnington {\em et al} \cite{ehp2}, and also recently confirmed by one of the authors (Tr{\" a}bert 2010, private communication). Therefore, although Pinnington {\em et al} \cite{ehp1} have listed four sets of lifetimes for the levels of the 3s$^2$3d and 3s3p$^2$ configurations corresponding to the free multi-exponential fitting, constrained multi-exponential fitting, cascade simulation (VNET) and arbitrarily normalised decay curve (ANDC), in Table 7 we have only included their `best' estimates which are based on the  cascade simulation and ANDC analysis. However, all four sets of lifetimes from the work of Pinnington {\em et al} \cite{ehp2} are listed in Table 7.

It is clear from Table 7 that the various statistical analyses of the measurements by Pinnington {\em et al} \cite{ehp2} yield lifetimes with a wide range of values differing by up to $\sim$50\%, particularly for the levels of the 3s$^2$3d and 3s3p$^2$ configurations. Since the $\tau$ values corresponding to the ANDC analysis include cascading and are considered to be more suitable for comparison with theory \cite{ehp1},\cite{ehp2}, we will mostly focus on those results. For the 40 levels included in Table 7, all values of $\tau$ from both the {\sc grasp} and {\sc fac} codes agree within 10\%, irrespective of the complexity of a calculation. This indicates that the lifetimes are stable, and (as for the energy levels) the effect of elaborate CI is not important in the determination of $\tau$. In general, the agreement between theory ({\sc grasp} and {\sc fac}) and measurements is within 10\% for a majority of levels, but there are differences of up to 50\% for some levels of the 3s3p3d configuration. For these the measurements appear to be overestimating the lifetimes, in comparison to both the present and earlier calculations. 

The MBPT results of Safronova {\em et al} \cite{uis} are generally the largest (by up to a factor of two), whereas those of Froese-Fischer {\em et al} \cite{cff1} from the MCHF code are slightly on the smaller side {\em except} for the highest five levels for which they are larger (by up to almost a factor of two). In fact, for the highest five levels, although the MCHF values of $\tau$ are in agreement with the earlier (uncertain) measurements of Pinnington {\em et al} \cite{ehp1}, they are the largest among all calculations. As may be seen from Table 1, the MCHF energies for these five levels are comparatively in better agreement with the experimental results, and our energies from the {\sc grasp} and {\sc fac} codes are larger by up to 2\%. However, the present lifetimes should be reliable because an adjustment of our theoretical energies with  measurements for the relevant transitions, which contribute to the lifetimes, alters the listed results by less than 10\%. Finally, the lifetimes calculated by Singh {\em et al} \cite{mm2} from the CIV3 code differ for many levels from all other calculated results by up to 30\% -- see, for example, levels 22--27.  Since we have performed a variety of calculations with increasing amount of CI, and with two different and independent atomic structure codes, we have confidence in our results, because for the lowest 40 levels of Ti X the effect of large CI on energy levels and lifetimes is insignificant. In conclusion, we may state that the lifetimes listed by Singh {\em et al} differ from other calculations by up to 30\% for many levels and hence may not be reliable. These differences in lifetimes directly arise from the corresponding differences in radiative rates discussed in section 3 and shown in Table 5.             

Differences between our results for $\tau$ and those of   Singh {\em et al} \cite{mm2}  are even larger for higher excited levels of Ti X. For illustration in Table 8 we compare the lifetimes for levels of the 3s3p4$\ell$ configurations. Singh {\em et al} have not listed $\tau$ for all the desired levels, and among the common levels the discrepancies are up to a factor of four for several, such as 18,19, 29 and 31--34. The maximum discrepancy is for  3s3p($^3$P)4d $^4$F$^o_{5/2}$  (33), where the major contribution to the lifetime is from 4  E1 transitions, namely 3s3p$^2$ $^4$P$_{3/2}$ -- 3s3p($^3$P)4d $^4$F$^o_{5/2}$ (4--165: A = 5.08$\times$10$^{9}$ s$^{-1}$), 3s3p$^2$ $^2$D$_{3/2}$ -- 3s3p($^3$P)4d $^4$F$^o_{5/2}$ (6--165: A = 6.85$\times$10$^{9}$ s$^{-1}$), 3s3p($^3$P)4p $^2$P$_{3/2}$ -- 3s3p($^3$P)4d $^4$F$^o_{5/2}$ (117--165: A = 1.10$\times$10$^{9}$ s$^{-1}$) and 3s3p($^3$P)4p $^4$D$_{5/2}$ -- 3s3p($^3$P)4d $^4$F$^o_{5/2}$ (119--165: A = 1.07$\times$10$^{9}$ s$^{-1}$) -- the level indices corresponding to those of Table 3 (see supplementary data, available online at stacks.iop.org/PhysScr/vol/number/mmedia). Unfortunately, A- values have not been listed/calculated by Singh {\em et al} \cite{mm1} for any of these transitions. However, if we {\em exclude} the contribution of these four transitions then $\Sigma$A$_{ji}$ = 3.6$\times$10$^{9}$ s$^{-1}$, or equivalently $\tau$ = 278 ps, which is much closer to the listed value of 230 ps by Singh {\em et al} \cite{mm1}. Therefore, not only  are the reported results of Singh {\em et al} \cite{mm1},\cite{mm2} inaccurate, but also incomplete.

\section{Conclusions}

In the present work, energy levels, radiative rates, oscillator strengths and line strengths for transitions among 345 fine-structure levels of Ti X are computed  using the fully relativistic {\sc grasp} code, and results reported for electric and magnetic dipole and quadrupole transitions. For calculating these parameters an extensive CI (with up to 1387 levels) has been included, which has been observed to be very significant, particularly for the accurate determination of energy levels. Furthermore, analogous calculations have been performed with the {\sc fac} code and with the inclusion of even larger CI with up to 12,139 levels, but the additional CI included does not appreciably affect the magnitude or orderings of the lowest 345 energy levels. Based on a variety of comparisons among different calculations, the reported energy levels are assessed to be accurate to better than 1\%. 

There is a paucity of measured energies for a majority of  the levels of Ti X. However, for the common levels there is no major discrepancy  with our calculations, although the orderings slightly differ in a few cases. Other theoretical energies are available from a variety of methods/codes, but primarily for the lowest 40 levels, for which most of the calculations are in agreement in both the magnitude as well as the ordering. However,  theoretical energies \cite{mm1} are available for a larger number of levels, up to 303. Although the calculations of Singh {\em et al} \cite{mm1}  included extensive CI, their energy levels are not as accurate as presented in this paper. Discrepancies are greater, up to an order of magnitude,  for the A- values between their data and the present calculations. As for the energy levels, extensive comparisons, based on a variety of calculations with the {\sc grasp} and {\sc fac} codes,  have been made for the A- values, and the accuracy of these is assessed to be  $\sim$ 20\% for a majority of the strong transitions. 

Lifetimes are also reported for all levels, but  measurements are available for only a few. For levels of the 3s$^2$3d and 3s3p$^2$ configurations, agreement between theory and measurements is within 10\%. However, differences are larger (up to 50\%) for the levels of the 3s3p3d configuration for which the measurements appear to be overestimated. Finally, calculations for energy levels and radiative rates have been performed for up to 12,139 levels of Ti X, but for brevity results have been reported for only the lowest 345 levels. However, a complete set of results for all calculated parameters can be obtained on request from one of the authors (K.Aggarwal@qub.ac.uk).

\section*{Acknowledgment}
KMA  is thankful to  AWE Aldermaston for financial support.   



\clearpage

\begin{flushleft}
{\bf Table 1.} Energies (Ryd) for the lowest 40 levels of Ti X. 
\end{flushleft}
\begin{tabular}{rllrrrrrrrrrr} \hline
 & & & & & & & &  \\
Index  & \multicolumn{2}{c}{Configuration/Level} & NIST &  GRASP1   & GRASP2   & FAC1    & FAC2     & MCHF     &  MPPT    & CIV3    \\
& & & & & & & &   \\ \hline
& & & & & & & &   \\
  1  &  3s$^2$3p       &  $^2$P$^o$$_{1/2}$  & 0.00000  &   0.00000 &  0.00000 & 0.00000 &  0.00000 & 0.00000  & 0.00000  &  0.0000 \\
  2  &  3s$^2$3p       &  $^2$P$^o$$_{3/2}$  & 0.06875  &   0.06804 &  0.06804 & 0.06779 &  0.06781 & 0.06652  & 0.06874  &  0.0689 \\
  3  &  3s3p$^2$       &  $^4$P$$$$$_{1/2}$  & 1.46175  &   1.44410 &  1.44344 & 1.44680 &  1.44797 & 1.44781  & 1.46178  &  1.4612 \\
  4  &  3s3p$^2$       &  $^4$P$$$$$_{3/2}$  & 1.48771  &   1.46969 &  1.46905 & 1.47229 &  1.47348 & 1.47324  & 1.48778  &  1.4848 \\
  5  &  3s3p$^2$       &  $^4$P$$$$$_{5/2}$  & 1.52463  &   1.50641 &  1.50576 & 1.50881 &  1.51000 & 1.50870  & 1.52473  &  1.5241 \\
  6  &  3s3p$^2$       &  $^2$D$$$$$_{3/2}$  & 1.93237  &   1.93842 &  1.93586 & 1.93636 &  1.93613 & 1.92770  & 1.93211  &  1.9347 \\
  7  &  3s3p$^2$       &  $^2$D$$$$$_{5/2}$  & 1.93743  &   1.94335 &  1.94080 & 1.94123 &  1.94100 & 1.93237  & 1.93712  &  1.9372 \\
  8  &  3s3p$^2$       &  $^2$S$$$$$_{1/2}$  & 2.40990  &   2.45830 &  2.45047 & 2.44751 &  2.44402 & 2.41723  & 2.41116  &  2.4091 \\
  9  &  3s3p$^2$       &  $^2$P$$$$$_{1/2}$  & 2.56113  &   2.61814 &  2.61001 & 2.60685 &  2.60515 & 2.58795  & 2.56132  &  2.5612 \\
 10  &  3s3p$^2$       &  $^2$P$$$$$_{3/2}$  & 2.59912  &   2.65634 &  2.64820 & 2.64493 &  2.64336 & 2.62591  & 2.59928  &  2.6075 \\
 11  &  3s$^2$3d       &  $^2$D$$$$$_{3/2}$  & 3.14674  &   3.20091 &  3.19816 & 3.18843 &  3.18453 &	       & 3.14762  &  3.1482 \\
 12  &  3s$^2$3d       &  $^2$D$$$$$_{5/2}$  & 3.15170  &   3.20520 &  3.20248 & 3.19254 &  3.18865 &	       & 3.15265  &  3.1551 \\
 13  &  3p$^3$         &  $^2$D$^o$$_{3/2}$  & 3.76715  &   3.76536 &  3.76543 & 3.76621 &  3.76612 & 3.77601  & 3.76680  &  3.7659 \\
 14  &  3p$^3$         &  $^2$D$^o$$_{5/2}$  & 3.77597  &   3.77364 &  3.77373 & 3.77445 &  3.77440 & 3.78481  & 3.77553  &  3.7705 \\
 15  &  3p$^3$         &  $^4$S$^o$$_{3/2}$  & 3.86116  &   3.88179 &  3.88267 & 3.88351 &  3.88365 & 3.87755  & 3.86076  &  3.8615 \\
 16  &  3s3p($^3$P)3d  &  $^4$F$^o$$_{3/2}$  &          &   4.23152 &  4.23011 & 4.22493 &  4.22242 & 4.22677  & 4.23042  &  4.2306 \\
 17  &  3s3p($^3$P)3d  &  $^4$F$^o$$_{5/2}$  & 4.24568  &   4.24580 &  4.24439 & 4.23912 &  4.23662 & 4.24080  & 4.24494  &  4.2458 \\
 18  &  3p$^3$         &  $^2$P$^o$$_{1/2}$  & 4.21135  &   4.24936 &  4.24850 & 4.24606 &  4.24205 & 4.22627  & 4.21173  &  4.2143 \\
 19  &  3p$^3$         &  $^2$P$^o$$_{3/2}$  & 4.21651  &   4.25371 &  4.25277 & 4.25032 &  4.24638 & 4.21328  & 4.21704  &  4.2089 \\
 20  &  3s3p($^3$P)3d  &  $^4$F$^o$$_{7/2}$  & 4.26659  &   4.26637 &  4.26498 & 4.25959 &  4.25709 & 4.26121  & 4.26587  &  4.2671 \\
 21  &  3s3p($^3$P)3d  &  $^4$F$^o$$_{9/2}$  & 4.29466  &   4.29413 &  4.29277 & 4.28723 &  4.28473 &	       & 4.29407  &  4.2943 \\
 22  &  3s3p($^3$P)3d  &  $^4$P$^o$$_{5/2}$  & 4.56977  &   4.58788 &  4.58402 & 4.57717 &  4.57617 & 4.57983  & 4.57006  &  4.5707 \\
 23  &  3s3p($^3$P)3d  &  $^4$P$^o$$_{3/2}$  & 4.58313  &   4.60249 &  4.59867 & 4.59181 &  4.59057 & 4.59266  & 4.62416  &  4.5860 \\
 24  &  3s3p($^3$P)3d  &  $^4$D$^o$$_{1/2}$  & 4.61875  &   4.61544 &  4.61167 & 4.60479 &  4.60319 & 4.60410  & 4.59581  &  4.6186 \\
 25  &  3s3p($^3$P)3d  &  $^4$P$^o$$_{1/2}$  & 4.59427  &   4.63804 &  4.63427 & 4.62732 &  4.62590 & 4.62599  & 4.61740  &  4.5952 \\
 26  &  3s3p($^3$P)3d  &  $^4$D$^o$$_{3/2}$  & 4.62461  &   4.64560 &  4.64190 & 4.63485 &  4.63305 & 4.63337  & 4.58384  &  4.6239 \\
 27  &  3s3p($^3$P)3d  &  $^4$D$^o$$_{5/2}$  & 4.62795  &   4.65036 &  4.64672 & 4.63957 &  4.63748 & 4.63842  & 4.62818  &  4.6328 \\
 28  &  3s3p($^3$P)3d  &  $^4$D$^o$$_{7/2}$  & 4.62755  &   4.65100 &  4.64740 & 4.64015 &  4.63777 & 4.63994  & 4.62806  &  4.6453 \\
 29  &  3s3p($^3$P)3d  &  $^2$D$^o$$_{3/2}$  & 4.72979  &   4.77285 &  4.76983 & 4.76231 &  4.75674 & 4.74696  & 4.73063  &  4.7191 \\
 30  &  3s3p($^3$P)3d  &  $^2$D$^o$$_{5/2}$  & 4.73051  &   4.77320 &  4.77015 & 4.76260 &  4.75705 & 4.74756  & 4.73128  &  4.7205 \\
 31  &  3s3p($^3$P)3d  &  $^2$F$^o$$_{5/2}$  & 4.94969  &   5.01652 &  5.01009 & 5.00050 &  4.98724 & 4.96233  & 4.95032  &  4.9416 \\
 32  &  3s3p($^3$P)3d  &  $^2$F$^o$$_{7/2}$  & 5.00420  &   5.07065 &  5.06429 & 5.05452 &  5.04120 & 5.01488  & 5.00493  &  4.9963 \\
 33  &  3s3p($^3$P)3d  &  $^2$P$^o$$_{3/2}$  & 5.38048  &   5.47927 &  5.46919 & 5.45778 &  5.44869 & 5.40594  & 5.38491  &  5.3893 \\
 34  &  3s3p($^3$P)3d  &  $^2$P$^o$$_{1/2}$  & 5.40519  &   5.50370 &  5.49302 & 5.48152 &  5.47274 & 5.42944  & 5.40974  &  5.4079 \\
 35  &  3s3p($^1$P)3d  &  $^2$F$^o$$_{7/2}$  & 5.42225  &   5.52008 &  5.51189 & 5.50165 &  5.49264 & 5.45541  & 5.42677  &  5.4545 \\
 36  &  3s3p($^1$P)3d  &  $^2$F$^o$$_{5/2}$  & 5.43543  &   5.53338 &  5.52518 & 5.51500 &  5.50590 & 5.46729  & 5.43964  &  5.4673 \\
 37  &  3s3p($^1$P)3d  &  $^2$P$^o$$_{1/2}$  & 5.58268  &   5.69669 &  5.68645 & 5.67545 &  5.65982 & 5.62072  & 5.58463  &  5.5832 \\
 38  &  3s3p($^1$P)3d  &  $^2$P$^o$$_{3/2}$  & 5.58836  &   5.70232 &  5.69149 & 5.68028 &  5.66493 & 5.62813  & 5.61988  &  5.5918 \\
 39  &  3s3p($^1$P)3d  &  $^2$D$^o$$_{3/2}$  & 5.61581  &   5.73235 &  5.72404 & 5.71159 &  5.69535 & 5.65955  & 5.59101  &  5.6394 \\
 40  &  3s3p($^1$P)3d  &  $^2$D$^o$$_{5/2}$  & 5.62423  &   5.74071 &  5.73236 & 5.71973 &  5.70343 & 5.66914  & 5.62889  &  5.6547 \\
& & & & & & & & \\ \hline            								                	 
\end{tabular}   								   					       
			      							   					       
\vspace*{0.9 cm}													       
\begin{flushleft}													       
{\small
NIST: http://www.nist.gov/pml/data/asd.cfm \\
GRASP1: present calculations from the {\sc grasp} code with 530 levels \\ 
GRASP2: present calculations from the {\sc grasp} code with 1387 levels\\ 
FAC1: present calculations from the {\sc fac} code with 1391 levels  \\  
FAC2: present calculations from the {\sc fac} code with 12,139 levels \\ 										
MCHF: Froese$-$Fischer {\em et al} \cite{cff1} \\ 
MPPT: Santana {\em et al} \cite{sit} \\ 
CIV3: Singh {\em et al} \cite{mm1} \\ 
															       
}															       
\end{flushleft} 

\clearpage
\begin{flushleft}
{\bf Table 2a.} Energies (Ryd) for the 3p$^2$3d levels of Ti X. 
\end{flushleft}
\begin{tabular}{rllrrrrrrrrrr} \hline
 & & & & & & & &  \\
Index  & \multicolumn{2}{c}{Configuration/Level} & NIST &  GRASP1 & GRASP2   & FAC1    & FAC2     & CIV3a    & CIV3b   \\
& & & & & & & &   \\ \hline
& & & & & & & &   \\
  1  &  3p$^2$($^1$D)3d  &  $^2$F$_{5/2}$  & 	      & 6.34665   & 6.33646  & 6.33059 & 6.32971  &	    & 6.2161  \\
  2  &  3p$^2$($^1$D)3d  &  $^2$F$_{7/2}$  & 	      & 6.37286   & 6.36272  & 6.35672 & 6.35590  &	    & 6.2461  \\
  3  &  3p$^2$($^3$P)3d  &  $^4$F$_{3/2}$  & 	      & 6.42998   & 6.41750  & 6.41113 & 6.40999  &	    & 6.4016  \\
  4  &  3p$^2$($^3$P)3d  &  $^4$F$_{5/2}$  & 6.41120  & 6.44515   & 6.43271  & 6.42628 & 6.42515  &	    & 6.4168  \\
  5  &  3p$^2$($^3$P)3d  &  $^4$F$_{7/2}$  & 6.43255  & 6.46591   & 6.45354  & 6.44701 & 6.44589  &	    & 6.4381  \\
  6  &  3p$^2$($^3$P)3d  &  $^4$F$_{9/2}$  & 6.45741  & 6.49081   & 6.47848  & 6.47184 & 6.47068  &	    & 6.4654  \\
  7  &  3p$^2$($^3$P)3d  &  $^2$P$_{3/2}$  & 6.50224  & 6.54153   & 6.52532  & 6.51848 & 6.51726  & 6.5440  & 6.5664  \\
  8  &  3p$^2$($^3$P)3d  &  $^4$D$_{1/2}$  & 	      & 6.56898   & 6.55211  & 6.54516 & 6.54372  &	    & 6.5512  \\
  9  &  3p$^2$($^3$P)3d  &  $^4$D$_{3/2}$  & 6.55630  & 6.60115   & 6.58358  & 6.57641 & 6.57478  &	    & 6.5568  \\
 10  &  3p$^2$($^3$P)3d  &  $^4$D$_{5/2}$  & 6.55972  & 6.60511   & 6.58743  & 6.58021 & 6.57853  &	    & 6.5660  \\
 11  &  3p$^2$($^1$D)3d  &  $^2$P$_{1/2}$  & 	      & 6.61641   & 6.59986  & 6.59271 & 6.59134  & 7.2282  & 6.6128  \\
 12  &  3p$^2$($^3$P)3d  &  $^4$D$_{7/2}$  & 6.57856  & 6.62307   & 6.60562  & 6.59832 & 6.59668  &	    & 6.5789  \\
 13  &  3p$^2$($^1$D)3d  &  $^2$G$_{7/2}$  & 	      & 6.74538   & 6.72504  & 6.71773 & 6.71240  &	    & 6.6321  \\
 14  &  3p$^2$($^1$D)3d  &  $^2$G$_{9/2}$  & 	      & 6.75488   & 6.73479  & 6.72737 & 6.72211  &	    & 6.6403  \\
 15  &  3p$^2$($^1$D)3d  &  $^2$D$_{5/2}$  & 6.85105  & 6.91187   & 6.89405  & 6.88534 & 6.88275  & 6.9020  & 6.8433  \\
 16  &  3p$^2$($^1$D)3d  &  $^2$D$_{3/2}$  & 6.86324  & 6.92282   & 6.90594  & 6.89727 & 6.89460  & 6.9081  & 6.8468  \\
 17  &  3p$^2$($^3$P)3d  &  $^4$P$_{5/2}$  & 6.90029  & 6.97370   & 6.94960  & 6.94088 & 6.93873  & 7.0046  & 6.8889  \\
 18  &  3p$^2$($^3$P)3d  &  $^4$P$_{3/2}$  & 6.90762  & 6.98324   & 6.95814  & 6.94946 & 6.94739  & 7.0157  & 6.9037  \\
 19  &  3p$^2$($^3$P)3d  &  $^4$P$_{1/2}$  & 6.91245  & 6.98948   & 6.96401  & 6.95535 & 6.95329  & 7.0227  & 6.9125  \\
 20  &  3p$^2$($^3$P)3d  &  $^2$P$_{1/2}$  & 	      & 7.24324   & 7.20424  & 7.19291 & 7.18667  & 6.5885  & 7.2691  \\
 21  &  3p$^2$($^1$D)3d  &  $^2$P$_{3/2}$  & 	      & 7.25439   & 7.21824  & 7.20670 & 7.20002  & 7.2336  & 7.2894  \\
 22  &  3p$^2$($^1$D)3d  &  $^2$S$_{1/2}$  & 	      & 7.30240   & 7.27855  & 7.27368 & 7.26385  & 7.2546  & 7.3487  \\
 23  &  3p$^2$($^1$S)3d  &  $^2$D$_{3/2}$  & 7.18756  & 7.30949   & 7.26709  & 7.25619 & 7.24776  & 7.2717  & 7.1846  \\
 24  &  3p$^2$($^1$S)3d  &  $^2$D$_{5/2}$  & 7.22562  & 7.34937   & 7.30693  & 7.29611 & 7.28713  & 7.3050  & 7.2372  \\
 25  &  3p$^2$($^3$P)3d  &  $^2$F$_{5/2}$  & 7.29581  & 7.41742   & 7.38672  & 7.37550 & 7.36768  &	    & 7.2979  \\
 26  &  3p$^2$($^3$P)3d  &  $^2$F$_{7/2}$  & 7.31161  & 7.43297   & 7.40275  & 7.39149 & 7.38365  &	    & 7.3117  \\
 27  &  3p$^2$($^3$P)3d  &  $^2$D$_{5/2}$  & 7.64013  & 7.80218   & 7.75977  & 7.74774 & 7.73701  & 7.7602  & 7.7123  \\
 28  &  3p$^2$($^3$P)3d  &  $^2$D$_{3/2}$  & 7.66253  & 7.82099   & 7.78091  & 7.76877 & 7.75823  & 7.7825  & 7.7360  \\
& & & & & & & & \\ \hline            								                	 
\end{tabular}   								   					       
			      							   					       
\begin{flushleft}													       
{\small
NIST: http://www.nist.gov/pml/data/asd.cfm \\
GRASP1: present calculations from the {\sc grasp} code with 530 levels \\ 
GRASP2: present calculations from the {\sc grasp} code with 1387 levels\\ 
FAC1: present calculations from the {\sc fac} code with 1391 levels  \\  
FAC2: present calculations from the {\sc fac} code with 12,139 levels \\ 										
CIV3a: Gupta and Msezane \cite{gm} \\ 
CIV3b: Singh {\em et al} \cite{mm1} \\ 
															       
}															       
\end{flushleft} 

\clearpage
\begin{flushleft}
{\bf Table 2b.} Energies (Ryd) for the 3s3d$^2$ levels of Ti X. 
\end{flushleft}
\begin{tabular}{rllrrrrrrrrrr} \hline
 & & & & & & & &  \\
Index  & \multicolumn{2}{c}{Configuration/Level} & NIST &  GRASP1 & GRASP2   & FAC1    & FAC2     & CIV3a    & CIV3b   \\
& & & & & & & &   \\ \hline
& & & & & & & &   \\
  1  &  3s3d$^2$  &  $^4$F$_{3/2}$  & 	       &  7.51027  & 7.49224  &  7.47401 & 7.47068 &	     & 7.5446  \\
  2  &  3s3d$^2$  &  $^4$F$_{5/2}$  & 	       &  7.51217  & 7.49413  &  7.47586 & 7.47253 &	     & 7.5474  \\
  3  &  3s3d$^2$  &  $^4$F$_{7/2}$  & 	       &  7.51479  & 7.49675  &  7.47840 & 7.47507 &	     & 7.5514  \\
  4  &  3s3d$^2$  &  $^4$F$_{9/2}$  & 	       &  7.51813  & 7.50008  &  7.48162 & 7.47830 &	     & 7.5566  \\
  5  &  3s3d$^2$  &  $^4$P$_{1/2}$  & 	       &  7.75351  & 7.73407  &  7.71566 & 7.71024 & 7.7555  & 7.7399  \\
  6  &  3s3d$^2$  &  $^4$P$_{3/2}$  & 	       &  7.75469  & 7.73526  &  7.71681 & 7.71139 & 7.7572  & 7.7416  \\
  7  &  3s3d$^2$  &  $^4$P$_{5/2}$  & 	       &  7.75613  & 7.73671  &  7.71820 & 7.71279 & 7.7588  & 7.7446  \\
  8  &  3s3d$^2$  &  $^2$G$_{7/2}$  & 7.93930  &  8.08280  & 8.04641  &  8.02619 & 8.01464 &	     & 7.9339  \\
  9  &  3s3d$^2$  &  $^2$G$_{9/2}$  & 7.93990  &  8.08365  & 8.04717  &  8.02699 & 8.01542 &	     & 7.9349  \\
 10  &  3s3d$^2$  &  $^2$D$_{5/2}$  & 	       &  8.10860  & 8.06208  &  8.04190 & 8.03081 & 8.0777  & 8.0537  \\
 11  &  3s3d$^2$  &  $^2$D$_{3/2}$  & 	       &  8.10993  & 8.06267  &  8.04255 & 8.03137 & 8.0780  & 8.0539  \\
 12  &  3s3d$^2$  &  $^2$F$_{5/2}$  & 8.28796  &  8.49278  & 8.42810  &  8.40291 & 8.38858 &	     & 8.2854  \\
 13  &  3s3d$^2$  &  $^2$F$_{7/2}$  & 	       &  8.49514  & 8.43050  &  8.40511 & 8.39081 &	     & 8.2899  \\
 14  &  3s3d$^2$  &  $^2$S$_{1/2}$  & 	       &  8.66262  & 8.60999  &  8.59064 & 8.56850 & 8.6203  & 8.7500  \\
 15  &  3s3d$^2$  &  $^2$P$_{1/2}$  & 	       &  8.73886  & 8.65433  &  8.63040 & 8.61479 & 8.7259  & 8.7519  \\
 16  &  3s3d$^2$  &  $^2$P$_{3/2}$  & 	       &  8.74447  & 8.65955  &  8.63554 & 8.61994 & 8.7324  & 8.7586  \\
& & & & & & & & \\ \hline            								                	 
\end{tabular}   								   					       
			      							   					       
\begin{flushleft}													       
{\small
NIST: http://www.nist.gov/pml/data/asd.cfm \\
GRASP1: present calculations from the {\sc grasp} code with 530 levels \\ 
GRASP2: present calculations from the {\sc grasp} code with 1387 levels\\ 
FAC1: present calculations from the {\sc fac} code with 1391 levels  \\  
FAC2: present calculations from the {\sc fac} code with 12,139 levels \\ 										
CIV3a: Gupta and Msezane \cite{gm} \\ 
CIV3b: Singh {\em et al} \cite{mm1} \\ 
															       
}															       
\end{flushleft} 

\clearpage
\begin{flushleft}
{\bf Table 2c.} Energies (Ryd) for the 3p3d$^2$ levels of Ti X. 
\end{flushleft}
{\small
\begin{tabular}{rllrrrrrrrrrr} \hline
 & & & & & & & &  \\
Index  & \multicolumn{2}{c}{Configuration/Level} & GRASP1 & GRASP2    & FAC1      & FAC2      & CIV3a    & CIV3b      \\
& & & & & & & &   \\ \hline
& & & & & & & &   \\
  1  &  3p3d$^2$($^3$F)  & $^4$G$^o_{5/2 }$   &  9.06644  &  9.05601  &  9.04014  &  9.03403  &          & 	      \\
  2  &  3p3d$^2$($^3$F)  & $^4$G$^o_{7/2 }$   &	 9.08278  &  9.07240  &  9.05648  &  9.05040  &          & 	      \\
  3  &  3p3d$^2$($^3$F)  & $^4$G$^o_{9/2 }$   &	 9.10405  &  9.09372  &  9.07775  &  9.07169  &          & 	      \\
  4  &  3p3d$^2$($^3$F)  & $^4$G$^o_{11/2}$   &  9.13074  &  9.12047  &  9.10445  &  9.09842  &          & 	      \\
  5  &  3p3d$^2$($^3$F)  & $^2$F$^o_{5/2 }$   &  9.29233  &  9.27857  &  9.26189  &  9.24731  & 10.0357  &  9.2944    \\
  6  &  3p3d$^2$($^3$F)  & $^2$F$^o_{7/2 }$   &  9.32685  &  9.31328  &  9.29648  &  9.28096  & 10.0389  &  9.3237    \\
  7  &  3p3d$^2$($^3$F)  & $^2$D$^o_{3/2 }$   &  9.33939  &  9.32655  &  9.30963  &  9.29808  &  9.3377  &  9.4314    \\
  8  &  3p3d$^2$($^3$F)  & $^2$D$^o_{5/2 }$   &	 9.35451  &  9.34159  &  9.32460  &  9.31288  &  9.3523  &  9.4402    \\
  9  &  3p3d$^2$($^3$P)  & $^4$D$^o_{1/2 }$   &  9.44018  &  9.42457  &  9.40642  &  9.40012  &  9.4391  &  9.4952    \\
 10  &  3p3d$^2$($^3$P)  & $^4$D$^o_{3/2 }$   &  9.45115  &  9.43570  &  9.41753  &  9.41073  &  9.4496  &  9.5045    \\
 11  &  3p3d$^2$($^3$P)  & $^2$S$^o_{1/2 }$   &	 9.46243  &  9.45027  &  9.43218  &  9.42780  &          &  9.5752    \\
 12  &  3p3d$^2$($^3$P)  & $^4$D$^o_{5/2 }$   &  9.46609  &  9.45090  &  9.43263  &  9.42500  &  9.4641  &  9.5194    \\
 13  &  3p3d$^2$($^3$P)  & $^4$D$^o_{7/2 }$   &	 9.48099  &  9.46609  &  9.44757  &  9.43872  &  9.4790  &  9.5388    \\
 14  &  3p3d$^2$($^3$F)  & $^4$F$^o_{3/2 }$   &	 9.51669  &  9.50631  &  9.48606  &  9.47067  &  9.5152  &  9.5960    \\
 15  &  3p3d$^2$($^3$F)  & $^4$F$^o_{5/2 }$   &  9.52197  &  9.51157  &  9.49127  &  9.47614  &  9.5207  &  9.6020    \\
 16  &  3p3d$^2$($^3$F)  & $^4$F$^o_{7/2 }$   &  9.52861  &  9.51769  &  9.49756  &  9.48234  &  9.5292  &  9.6105    \\
 17  &  3p3d$^2$($^3$F)  & $^4$F$^o_{9/2 }$   &  9.52992  &  9.51796  &  9.49816  &  9.48123  &  9.5379  &  9.6214    \\
 18  &  3p3d$^2$($^1$G)  & $^2$G$^o_{7/2 }$   &  9.55830  &  9.54209  &  9.52379  &  9.50335  &          &  9.6501    \\
 19  &  3p3d$^2$($^1$G)  & $^2$G$^o_{9/2 }$   &  9.56344  &  9.54807  &  9.52919  &  9.50988  &          &  9.6559    \\
 20  &  3p3d$^2$($^3$P)  & $^4$P$^o_{1/2 }$   &	 9.58527  &  9.56896  &  9.54984  &  9.54361  &  9.5721  &  9.6914    \\
 21  &  3p3d$^2$($^3$P)  & $^4$P$^o_{3/2 }$   &	 9.58765  &  9.57096  &  9.55176  &  9.54515  &  9.5829  &  9.7041    \\
 22  &  3p3d$^2$($^3$P)  & $^4$P$^o_{5/2 }$   &	 9.61013  &  9.59354  &  9.57429  &  9.56775  &  9.6043  &  9.7250    \\
 23  &  3p3d$^2$($^1$G)  & $^2$H$^o_{9/2 }$   &  9.68765  &  9.66080  &  9.64204  &  9.62000  &          & 	      \\
 24  &  3p3d$^2$($^1$G)  & $^2$H$^o_{11/2}$   &  9.73277  &  9.70548  &  9.68663  &  9.66441  &          & 	      \\
 25  &  3p3d$^2$($^3$F)  & $^4$D$^o_{7/2 }$   &  9.77877  &  9.76766  &  9.74534  &  9.72206  &  9.7764  &  9.8505    \\
 26  &  3p3d$^2$($^3$F)  & $^4$D$^o_{5/2 }$   &  9.78280  &  9.77182  &  9.74944  &  9.72521  &  9.7792  &  9.8529    \\
 27  &  3p3d$^2$($^3$F)  & $^4$D$^o_{3/2 }$   &  9.78582  &  9.77497  &  9.75258  &  9.72778  &  9.7812  &  9.8557    \\
 28  &  3p3d$^2$($^3$F)  & $^4$D$^o_{1/2 }$   &  9.78793  &  9.77715  &  9.75475  &  9.72965  &  9.7826  &  9.8578    \\
 29  &  3p3d$^2$($^1$D)  & $^2$P$^o_{3/2 }$   &	 9.87795  &  9.85669  &  9.83576  &  9.81751  &  9.8551  &  9.9966    \\
 30  &  3p3d$^2$($^1$D)  & $^2$P$^o_{1/2 }$   &	 9.90070  &  9.88082  &  9.85986  &  9.84219  &  9.8792  & 10.0205    \\
 31  &  3p3d$^2$($^3$P)  & $^4$S$^o_{3/2 }$   &	10.05933  & 10.04611  & 10.02388  &  9.97579  & 10.0512  & 10.2028    \\
 32  &  3p3d$^2$($^1$D)  & $^2$F$^o_{5/2 }$   & 10.06733  & 10.04026  & 10.01672  &  9.98816  &  9.2849  & 10.0711    \\
 33  &  3p3d$^2$($^1$D)  & $^2$F$^o_{7/2 }$   & 10.07193  & 10.04469  & 10.02127  &  9.99193  &  9.3190  & 10.0745    \\
 34  &  3p3d$^2$($^1$D)  & $^2$D$^o_{3/2 }$   & 10.22278  & 10.20270  & 10.17722  & 10.12367  & 10.2065  & 10.3196    \\
 35  &  3p3d$^2$($^1$D)  & $^2$D$^o_{5/2 }$   & 10.24060  & 10.22105  & 10.19540  & 10.14128  & 10.2260  & 10.3386    \\
 36  &  3p3d$^2$($^1$S)  & $^2$P$^o_{1/2 }$   &	10.25996  & 10.22367  & 10.20032  & 10.17661  & 10.2186  & 10.3706    \\
 37  &  3p3d$^2$($^1$S)  & $^2$P$^o_{3/2 }$   & 10.30470  & 10.26935  & 10.24568  & 10.22066  & 10.2641  & 10.4152    \\
 38  &  3p3d$^2$($^1$G)  & $^2$F$^o_{7/2 }$   & 10.39127  & 10.36476  & 10.33820  & 10.27666  & 10.3702  & 10.4105    \\
 39  &  3p3d$^2$($^1$G)  & $^2$F$^o_{5/2 }$   &	10.41226  & 10.38705  & 10.36062  & 10.29641  & 10.3834  & 10.4251    \\
 40  &  3p3d$^2$($^3$F)  & $^2$G$^o_{9/2 }$   & 10.44495  & 10.41080  & 10.38336  & 10.33165  &          & 10.5408    \\
 41  &  3p3d$^2$($^3$F)  & $^2$G$^o_{7/2 }$   &	10.45609  & 10.42297  & 10.39545  & 10.34262  &          & 10.5452    \\
 42  &  3p3d$^2$($^3$P)  & $^2$D$^o_{5/2 }$   &	10.75154  & 10.71304  & 10.68426  & 10.61207  & 10.7189  & 10.8469    \\
 43  &  3p3d$^2$($^3$P)  & $^2$D$^o_{3/2 }$   & 10.75566  & 10.71684  & 10.68799  & 10.61648  & 10.7220  & 10.8511    \\
 44  &  3p3d$^2$($^3$P)  & $^2$P$^o_{1/2 }$   & 10.91107  & 10.87435  & 10.84730  & 10.76004  & 10.8455  & 11.0307    \\
 45  &  3p3d$^2$($^3$P)  & $^2$P$^o_{3/2 }$   & 10.91659  & 10.87961  & 10.85279  & 10.76629  & 10.8508  & 11.0354    \\
& & & & & & & & \\ \hline            					        					
\end{tabular}   							        					      
}			      						        					      
\newpage						        					      
\begin{flushleft}							        					      
{\small 								      
NIST: http://www.nist.gov/pml/data/asd.cfm \\				      
GRASP1: present calculations from the {\sc grasp} code with 530 levels \\     
GRASP2: present calculations from the {\sc grasp} code with 1387 levels\\     
FAC1: present calculations from the {\sc fac} code with 1391 levels  \\       
FAC2: present calculations from the {\sc fac} code with 12,139 levels \\         								       
CIV3a: Gupta and Msezane \cite{gm} \\ 
CIV3b: Singh {\em et al} \cite{mm1} \\ 
															       
}															       
\end{flushleft} 

\clearpage
\begin{flushleft}
{\bf Table 2d.} Energies (Ryd) for the 3d$^3$ levels of Ti X. 
\end{flushleft}
\begin{tabular}{rllrrrrrrrrrr} \hline
 & & & & & & & &  \\
Index  & \multicolumn{2}{c}{Configuration/Level} & GRASP1 & GRASP2    & FAC1      & FAC2   & CIV3       \\
& & & & & & & &   \\ \hline
& & & & & & & &   \\
  1  &  3d$^3$($^4$F)    &  $^4$F$_{3/2 }$  &  12.36655  & 12.34171   &  12.29955  & 12.29263	&  12.4844    \\
  2  &  3d$^3$($^4$F)    &  $^4$F$_{5/2 }$  &  12.36840  & 12.34359   &  12.30143  & 12.29451	&  12.4874    \\
  3  &  3d$^3$($^4$F)    &  $^4$F$_{7/2 }$  &  12.37090  & 12.34613   &  12.30395  & 12.29704	&  12.4917    \\
  4  &  3d$^3$($^4$F)    &  $^4$F$_{9/2 }$  &  12.37396  & 12.34922   &  12.30702  & 12.30011	&  12.4972    \\
  5  &  3d$^3$($^2$G)    &  $^2$G$_{7/2 }$  &  12.60763  & 12.53373   &  12.48627  & 12.47270	&  12.6319    \\
  6  &  3d$^3$($^2$G)    &  $^2$G$_{9/2 }$  &  12.61019  & 12.53603   &  12.48844  & 12.47482	&  12.6369    \\
  7  &  3d$^3$($^2$H)    &  $^2$H$_{9/2 }$  &  12.64880  & 12.55926   &  12.50818  & 12.49339	&  	      \\
  8  &  3d$^3$($^2$H)    &  $^2$H$_{11/2}$  &  12.65055  & 12.56067   &  12.50949  & 12.49461	&  	      \\
  9  &  3d$^3$($^4$P)    &  $^4$P$_{1/2 }$  &  12.60952  & 12.58672   &  12.53994  & 12.53041	&  12.6831    \\
 10  &  3d$^3$($^4$P)    &  $^4$P$_{3/2 }$  &  12.61069  & 12.58786   &  12.54111  & 12.53157	&  12.6849    \\
 11  &  3d$^3$($^4$P)    &  $^4$P$_{5/2 }$  &  12.61269  & 12.58998   &  12.54325  & 12.53371	&  12.6879    \\
 12  &  3d$^3$($^2$D 3)  &  $^2$D$_{3/2 }$  &  12.76502  & 12.69174   &  12.65068  & 12.63496	&  12.7901    \\
 13  &  3d$^3$($^2$D 3)  &  $^2$D$_{5/2 }$  &  12.78396  & 12.71753   &  12.65451  & 12.63898	&  12.7947    \\
 14  &  3d$^3$($^2$P)    &  $^2$P$_{1/2 }$  &  12.78193  & 12.71933   &  12.66891  & 12.65382	&  12.9125    \\
 15  &  3d$^3$($^2$P)    &  $^2$P$_{3/2 }$  &  12.78356  & 12.72243   &  12.67165  & 12.65597	&  12.9488    \\
 16  &  3d$^3$($^2$F)    &  $^2$F$_{7/2 }$  &  12.94071  & 12.83632   &  12.78407  & 12.76582	&  12.7607    \\
 17  &  3d$^3$($^2$F)    &  $^2$F$_{5/2 }$  &  12.93413  & 12.84344   &  12.78588  & 12.76504	&  12.7628    \\
 18  &  3d$^3$($^2$D 1)  &  $^2$D$_{5/2 }$  &  13.31196  & 13.20575   &  13.14516  & 13.11590	&  	      \\
 19  &  3d$^3$($^2$D 1)  &  $^2$D$_{3/2 }$  &  13.31313  & 13.20688   &  13.14631  & 13.11693	&  	      \\
& & & & & & & & \\ \hline            					        					
\end{tabular}   							        					      
			      						        					      
\begin{flushleft}							        					      
{\small 								      
NIST: http://www.nist.gov/pml/data/asd.cfm \\				      
GRASP1: present calculations from the {\sc grasp} code with 530 levels \\     
GRASP2: present calculations from the {\sc grasp} code with 1387 levels\\     
FAC1: present calculations from the {\sc fac} code with 1391 levels  \\       
FAC2: present calculations from the {\sc fac} code with 12,139 levels \\         								       
CIV3: Singh {\em et al} \cite{mm1} \\ 
															       
}															       
\end{flushleft} 

\clearpage
\begin{flushleft}
{\bf Table 2e.} Energies (Ryd) for the 3s3p4$\ell$ levels of Ti X. 
\end{flushleft}
{\small
\begin{tabular}{rllrrrrrrrrrr} \hline
Index  & \multicolumn{2}{c}{Configuration/Level} & GRASP1 & GRASP2    & FAC1      & FAC2      & CIV3a    & CIV3b      \\
 \hline
   1  &  3s3p($^3$P)4s  & $^4$P$^o_{ 1/2 }$   &  8.81616  &  8.79699  &  8.79837  &  8.79380  &  8.8065  &  8.8279   \\
   2  &  3s3p($^3$P)4s  & $^4$P$^o_{ 3/2 }$   &  8.83864  &  8.81933  &  8.82059  &  8.81584  &  8.8294  &  8.8512   \\
   3  &  3s3p($^3$P)4s  & $^4$P$^o_{ 5/2 }$   &  8.88162  &  8.86239  &  8.86323  &  8.85835  &  8.8727  &  8.8901   \\
   4  &  3s3p($^3$P)4s  & $^2$P$^o_{ 1/2 }$   &  9.00583  &  8.98517  &  8.99855  &  8.97871  &  8.9997  &  9.0026   \\
   5  &  3s3p($^3$P)4s  & $^2$P$^o_{ 3/2 }$   &  9.05284  &  9.03194  &  9.04458  &  9.02486  &  9.0462  &  9.0472   \\
   6  &  3s3p($^1$P)4s  & $^2$P$^o_{ 1/2 }$   &  9.80942  &  9.67113  &  9.66524  &  9.65582  &  9.6838  &  9.8031   \\
   7  &  3s3p($^1$P)4s  & $^2$P$^o_{ 3/2 }$   &  9.81104  &  9.67471  &  9.66919  &  9.65927  &  9.6870  &  9.8052   \\
   8  &  3s3p($^3$P)4p  & $^4$D$  _{ 1/2 }$   &  9.51019  &  9.49014  &  9.49309  &  9.48595  & 	 &  9.4829   \\
   9  &  3s3p($^3$P)4p  & $^4$D$  _{ 3/2 }$   &  9.53138  &  9.51114  &  9.51392  &  9.50675  & 	 &  9.4985   \\
  10  &  3s3p($^3$P)4p  & $^2$P$  _{ 1/2 }$   &  9.55527  &  9.53489  &  9.53774  &  9.53002  & 	 &  9.5276   \\
  11  &  3s3p($^3$P)4p  & $^2$P$  _{ 3/2 }$   &  9.56259  &  9.54197  &  9.54501  &  9.53729  & 	 &  9.5446   \\
  12  &  3s3p($^3$P)4p  & $^4$D$  _{ 5/2 }$   &  9.56915  &  9.55096  &  9.55329  &  9.54784  & 	 &  9.5244   \\
  13  &  3s3p($^3$P)4p  & $^4$D$  _{ 7/2 }$   &  9.60890  &  9.59075  &  9.59235  &  9.58704  & 	 &  9.5607   \\
  14  &  3s3p($^3$P)4p  & $^4$P$  _{ 1/2 }$   &  9.63407  &  9.61458  &  9.63288  &  9.61660  & 	 &  9.5415   \\
  15  &  3s3p($^3$P)4p  & $^4$P$  _{ 3/2 }$   &  9.64759  &  9.62856  &  9.64430  &  9.62990  & 	 &  9.5568   \\
  16  &  3s3p($^3$P)4p  & $^4$P$  _{ 5/2 }$   &  9.67809  &  9.65855  &  9.67639  &  9.65974  & 	 &  9.5823   \\
  17  &  3s3p($^3$P)4p  & $^4$S$  _{ 3/2 }$   &  9.69597  &  9.68049  &  9.68499  &  9.67938  & 	 &  9.7228   \\
  18  &  3s3p($^3$P)4p  & $^2$D$  _{ 3/2 }$   &  9.74115  &  9.71886  &  9.72903  &  9.71336  & 	 &  9.6393   \\
  19  &  3s3p($^3$P)4p  & $^2$D$  _{ 5/2 }$   &  9.77953  &  9.75828  &  9.76900  &  9.75265  & 	 &  9.6778   \\
  20  &  3s3p($^3$P)4p  & $^2$S$  _{ 1/2 }$   &  9.94237  &  9.92534  &  9.93595  &  9.91671  & 	 &  9.9583   \\
  21  &  3s3p($^1$P)4p  & $^2$P$  _{ 1/2 }$   & 10.53016  & 10.39466  & 10.39243  & 10.38219  & 	 & 10.4101   \\
  22  &  3s3p($^1$P)4p  & $^2$D$  _{ 3/2 }$   & 10.53019  & 10.39720  & 10.39589  & 10.38481  & 	 & 10.3205   \\
  23  &  3s3p($^1$P)4p  & $^2$D$  _{ 5/2 }$   & 10.54733  & 10.41466  & 10.41302  & 10.40182  & 	 & 10.3310   \\
  24  &  3s3p($^1$P)4p  & $^2$P$  _{ 3/2 }$   & 10.55082  & 10.41684  & 10.41494  & 10.40426  & 	 & 10.4250   \\
  25  &  3s3p($^1$P)4p  & $^2$S$  _{ 1/2 }$   & 10.61085  & 10.49166  & 10.49132  & 10.46684  & 	 & 10.5276   \\
  26  &  3s3p($^3$P)4d  & $^4$D$^o_{ 1/2 }$   & 10.53865  & 10.51649  & 10.51906  & 10.50836  & 	 & 10.5071   \\
  27  &  3s3p($^3$P)4d  & $^4$D$^o_{ 3/2 }$   & 10.54010  & 10.51814  & 10.52069  & 10.51007  & 	 & 10.5121   \\
  28  &  3s3p($^3$P)4d  & $^4$D$^o_{ 5/2 }$   & 10.54426  & 10.52213  & 10.52457  & 10.51387  & 	 & 10.5200   \\
  29  &  3s3p($^3$P)4d  & $^2$D$^o_{ 3/2 }$   & 10.54852  & 10.52430  & 10.52673  & 10.51474  & 	 & 10.5507   \\
  30  &  3s3p($^3$P)4d  & $^4$D$^o_{ 7/2 }$   & 10.55945  & 10.53721  & 10.53959  & 10.52891  & 	 & 10.5313   \\
  31  &  3s3p($^3$P)4d  & $^2$D$^o_{ 5/2 }$   & 10.56277  & 10.53835  & 10.54061  & 10.52858  & 	 & 10.5637   \\
  32  &  3s3p($^3$P)4d  & $^4$F$^o_{ 3/2 }$   & 10.61455  & 10.59621  & 10.60048  & 10.59106  & 	 & 10.5967   \\
  33  &  3s3p($^3$P)4d  & $^4$F$^o_{ 5/2 }$   & 10.62645  & 10.60737  & 10.61123  & 10.60160  & 	 & 10.6105   \\
  34  &  3s3p($^3$P)4d  & $^4$F$^o_{ 7/2 }$   & 10.64631  & 10.62768  & 10.63153  & 10.62198  & 	 & 10.6299   \\
  35  &  3s3p($^3$P)4d  & $^4$F$^o_{ 9/2 }$   & 10.66910  & 10.65101  & 10.65500  & 10.64553  & 	 & 10.6547   \\
  36  &  3s3p($^3$P)4d  & $^2$F$^o_{ 5/2 }$   & 10.67788  & 10.64962  & 10.65130  & 10.63986  & 	 & 10.5565   \\
  37  &  3s3p($^3$P)4d  & $^4$P$^o_{ 5/2 }$   & 10.68833  & 10.67230  & 10.67552  & 10.66600  & 	 & 10.7107   \\
  38  &  3s3p($^3$P)4d  & $^4$P$^o_{ 3/2 }$   & 10.70118  & 10.68579  & 10.68918  & 10.67922  & 	 & 10.7275   \\
  39  &  3s3p($^3$P)4d  & $^4$P$^o_{ 1/2 }$   & 10.71152  & 10.69634  & 10.69965  & 10.68984  & 	 & 10.7376   \\
  40  &  3s3p($^3$P)4d  & $^2$F$^o_{ 7/2 }$   & 10.71221  & 10.68340  & 10.68484  & 10.67263  & 	 & 10.5912   \\
  41  &  3s3p($^3$P)4d  & $^2$P$^o_{ 3/2 }$   & 10.74243  & 10.72679  & 10.72885  & 10.71500  & 	 & 10.7583   \\
  42  &  3s3p($^3$P)4d  & $^2$P$^o_{ 1/2 }$   & 10.76304  & 10.74760  & 10.74960  & 10.73482  & 	 & 10.7814   \\
  43  &  3s3p($^1$P)4d  & $^2$F$^o_{ 5/2 }$   & 11.46593  & 11.34003  & 11.33232  & 11.32047  & 	 & 11.3305   \\
  44  &  3s3p($^1$P)4d  & $^2$F$^o_{ 7/2 }$   & 11.47096  & 11.34375  & 11.33586  & 11.32425  & 	 & 11.3353   \\
  45  &  3s3p($^1$P)4d  & $^2$D$^o_{ 3/2 }$   & 11.48304  & 11.34811  & 11.34023  & 11.33136  & 	 & 11.4673   \\
  46  &  3s3p($^1$P)4d  & $^2$D$^o_{ 5/2 }$   & 11.48561  & 11.35076  & 11.34284  & 11.33391  & 	 & 11.4706   \\
  47  &  3s3p($^1$P)4d  & $^2$P$^o_{ 1/2 }$   & 11.61087  & 11.48817  & 11.48124  & 11.44921  & 	 & 11.6097   \\
  48  &  3s3p($^1$P)4d  & $^2$P$^o_{ 3/2 }$   & 11.61532  & 11.49341  & 11.48653  & 11.45369  &	         & 11.6142   \\
 \hline            					        					
\end{tabular} 
}
\newpage
\begin{tabular}{rllrrrrrrrrrr} \hline
 & & & & & & & &  \\
Index  & \multicolumn{2}{c}{Configuration/Level} & GRASP1 & GRASP2    & FAC1      & FAC2      & CIV3a    & CIV3b      \\
& & & & & & & &   \\ \hline
& & & & & & & &   \\
  49  &  3s3p($^3$P)4f  & $^4$G$  _{ 5/2 }$   & 11.06358  & 11.03971  & 11.05088  & 11.03380  &    &	       \\
  50  &  3s3p($^3$P)4f  & $^4$F$  _{ 7/2 }$   & 11.06899  & 11.04424  & 11.05658  & 11.03926  &    & 11.0508   \\
  51  &  3s3p($^3$P)4f  & $^4$F$  _{ 3/2 }$   & 11.08180  & 11.05398  & 11.06592  & 11.04788  &    & 11.0430   \\
  52  &  3s3p($^3$P)4f  & $^4$F$  _{ 9/2 }$   & 11.08282  & 11.05727  & 11.06971  & 11.05201  &    & 11.0567   \\
  53  &  3s3p($^3$P)4f  & $^4$F$  _{ 5/2 }$   & 11.09182  & 11.06568  & 11.07744  & 11.05979  &    & 11.0463   \\
  54  &  3s3p($^3$P)4f  & $^4$G$  _{ 7/2 }$   & 11.10490  & 11.07899  & 11.09070  & 11.07327  &    &	       \\
  55  &  3s3p($^3$P)4f  & $^4$G$  _{ 9/2 }$   & 11.12583  & 11.10220  & 11.11362  & 11.09628  &    &	       \\
  56  &  3s3p($^3$P)4f  & $^4$G$  _{ 11/2}$   & 11.13231  & 11.11115  & 11.12176  & 11.10492  &    &	       \\
  57  &  3s3p($^3$P)4f  & $^2$F$  _{ 5/2 }$   & 11.14392  & 11.10694  & 11.11865  & 11.09917  &    & 10.8565   \\
  58  &  3s3p($^3$P)4f  & $^2$F$  _{ 7/2 }$   & 11.14750  & 11.11155  & 11.12351  & 11.10421  &    & 10.8641   \\
  59  &  3s3p($^3$P)4f  & $^4$D$  _{ 7/2 }$   & 11.19049  & 11.16682  & 11.17855  & 11.16430  &    & 11.1370   \\
  60  &  3s3p($^3$P)4f  & $^4$D$  _{ 5/2 }$   & 11.20324  & 11.18065  & 11.19226  & 11.17826  &    & 11.1550   \\
  61  &  3s3p($^3$P)4f  & $^4$D$  _{ 3/2 }$   & 11.21380  & 11.19142  & 11.20333  & 11.18942  &    & 11.1679   \\
  62  &  3s3p($^3$P)4f  & $^4$D$  _{ 1/2 }$   & 11.22012  & 11.19785  & 11.20999  & 11.19613  &    & 11.1756   \\
  63  &  3s3p($^3$P)4f  & $^2$G$  _{ 7/2 }$   & 11.30703  & 11.28001  & 11.29242  & 11.26576  &    & 11.1990   \\
  64  &  3s3p($^3$P)4f  & $^2$G$  _{ 9/2 }$   & 11.34521  & 11.31916  & 11.33147  & 11.30415  &    & 11.2386   \\
  65  &  3s3p($^3$P)4f  & $^2$D$  _{ 5/2 }$   & 11.35018  & 11.32607  & 11.33787  & 11.31541  &    & 11.2452   \\
  66  &  3s3p($^3$P)4f  & $^2$D$  _{ 3/2 }$   & 11.37595  & 11.35233  & 11.36396  & 11.34115  &    & 11.2714   \\
  67  &  3s3p($^1$P)4f  & $^2$F$  _{ 5/2 }$   & 12.02486  & 11.88436  & 11.88503  & 11.86535  &    & 11.6168   \\
  68  &  3s3p($^1$P)4f  & $^2$F$  _{ 7/2 }$   & 12.02686  & 11.88650  & 11.88699  & 11.86726  &    & 11.6182   \\
  69  &  3s3p($^1$P)4f  & $^2$D$  _{ 3/2 }$   & 12.18970  & 12.02383  & 12.03302  & 11.99261  &    & 11.9323   \\
  70  &  3s3p($^1$P)4f  & $^2$D$  _{ 5/2 }$   & 12.19103  & 12.02573  & 12.03581  & 11.99387  &    & 11.9337   \\
  71  &  3s3p($^1$P)4f  & $^2$G$  _{ 9/2 }$   & 12.25872  & 11.78861  & 11.78167  & 11.72743  &    & 11.8532   \\
  72  &  3s3p($^1$P)4f  & $^2$G$  _{ 7/2 }$   & 12.26056  & 11.79082  & 11.78410  & 11.72886  &    & 11.8579   \\
& & & & & & & & \\ \hline            					        					
\end{tabular}   							        					      
			      						        					      
\begin{flushleft}							        					      
{\small 								      
GRASP1: present calculations from the {\sc grasp} code with 530 levels \\     
GRASP2: present calculations from the {\sc grasp} code with 1387 levels\\     
FAC1: present calculations from the {\sc fac} code with 1391 levels  \\       
FAC2: present calculations from the {\sc fac} code with 12,139 levels \\         								       
CIV3a: Gupta and Msezane \cite{gm} \\ 
CIV3b: Singh {\em et al} \cite{mm1} \\ 
															       
}															       
\end{flushleft} 

\clearpage
\begin{flushleft}
{\bf Table 5.} Comparison of radiative rates (A-values, s$^{-1}$) for transitions among the lowest 40 levels of Ti X. $a{\pm}b \equiv a{\times}$10$^{{\pm}b}$.
\end{flushleft}
\begin{tabular}{rrrrrrrrrr} \hline
I & J   &        f (GRASP2)  & GRASP1     & GRASP2      & FAC1        & FAC2        &  MCHF	  &  CIV3     & NIST	    \\
 \hline
    1  &   6   &  7.6$-$02   &  1.1$+$09  &  1.2$+$09   &   1.1$+$09  &  1.2$+$09  & 1.2$+$09	&   9.0$+$08  &   1.1$+$08  \\
    1  &   8   &  1.5$-$01   &  7.3$+$09  &  7.3$+$09   &   7.2$+$09  &  7.2$+$09  & 6.6$+$09	&   3.1$+$09  &   6.9$+$09  \\
    1  &   9   &  2.6$-$01   &  1.4$+$10  &  1.4$+$10   &   1.4$+$10  &  1.4$+$10  & 1.4$+$10	&   1.8$+$10  &   1.3$+$10  \\
    1  &  10   &  1.8$-$01   &  5.1$+$09  &  5.0$+$09   &   5.0$+$09  &  5.0$+$09  & 4.9$+$09	&   4.8$+$09  &   4.7$+$09  \\
    1  &  11   &  6.5$-$01   &  2.7$+$10  &  2.7$+$10   &   2.6$+$10  &  2.6$+$10  &	     	&   2.7$+$10  &   2.5$+$10  \\
    2  &   6   &  3.5$-$03   &  1.0$+$08  &  9.9$+$07   &   9.9$+$07  &  9.9$+$07  & 1.1$+$08	&   1.6$+$08  &   9.5$+$07  \\
    2  &   7   &  5.9$-$02   &  1.1$+$09  &  1.1$+$09   &   1.1$+$09  &  1.1$+$09  & 1.2$+$09	&   9.8$+$08  &   1.1$+$09  \\
    2  &   8   &  3.5$-$02   &  3.2$+$09  &  3.2$+$09   &   3.1$+$09  &  3.2$+$09  & 3.3$+$09	&   5.7$+$09  &   2.7$+$09  \\
    2  &   9   &  1.2$-$01   &  1.2$+$10  &  1.2$+$10   &   1.2$+$10  &  1.2$+$10  & 1.2$+$10	&   8.3$+$09  &   1.2$+$10  \\
    2  &  10   &  4.3$-$01   &  2.3$+$10  &  2.3$+$10   &   2.3$+$10  &  2.3$+$10  & 2.3$+$10	&   2.2$+$10  &   2.2$+$10  \\
    2  &  11   &  7.2$-$02   &  5.7$+$09  &  5.6$+$09   &   5.6$+$09  &  5.6$+$09  &	     	&   5.0$+$09  &   5.3$+$09  \\
    2  &  12   &  5.9$-$01   &  3.1$+$10  &  3.1$+$10   &   3.0$+$10  &  3.0$+$10  &	     	&   3.0$+$10  &   2.9$+$10  \\
    3  &  15   &  1.8$-$01   &  4.3$+$09  &  4.3$+$09   &   4.3$+$09  &  4.3$+$09  & 4.2$+$09	&   4.1$+$09  &   4.1$+$09  \\
    3  &  24   &  3.0$-$01   &  2.5$+$10  &  2.5$+$10   &   2.4$+$10  &  2.5$+$10  & 2.5$+$10	&   3.1$+$09  &   2.3$+$10  \\
    3  &  25   &  6.5$-$02   &  5.2$+$09  &  5.3$+$09   &   5.1$+$09  &  4.3$+$09  & 4.1$+$09	&   2.7$+$10  &   7.6$+$09  \\
    4  &  15   &  1.8$-$01   &  8.2$+$09  &  8.3$+$09   &   8.2$+$09  &  8.2$+$09  & 8.2$+$09	&   7.9$+$09  &   7.7$+$09  \\
    4  &  24   &  7.3$-$03   &  1.1$+$09  &  1.2$+$09   &   1.1$+$09  &  7.6$+$08  & 7.1$+$08	&   1.5$+$09  &   1.6$+$09  \\
    4  &  25   &  1.2$-$01   &  2.0$+$10  &  2.0$+$10   &   2.0$+$10  &  2.0$+$10  & 2.0$+$10	&   5.2$+$09  &   1.8$+$10  \\
    4  &  27   &  2.1$-$01   &  1.1$+$10  &  1.1$+$10   &   1.1$+$10  &  1.1$+$10  & 1.1$+$10	&   2.2$+$10  &   1.1$+$10  \\
    5  &  15   &  1.7$-$01   &  1.2$+$10  &  1.2$+$10   &   1.2$+$10  &  1.2$+$10  & 1.2$+$10	&   1.1$+$10  &   1.1$+$10  \\
    5  &  28   &  5.2$-$01   &  3.1$+$10  &  3.1$+$10   &   3.1$+$10  &  3.0$+$10  & 3.1$+$10	&   3.1$+$10  &   3.0$+$10  \\
    6  &  13   &  5.2$-$02   &  1.4$+$09  &  1.4$+$09   &   1.4$+$09  &  1.4$+$09  & 1.4$+$09	&   1.7$+$09  &   1.6$+$09  \\
    6  &  14   &  7.9$-$03   &  1.4$+$08  &  1.4$+$08   &   1.4$+$08  &  1.4$+$08  & 1.6$+$08	&   1.3$+$08  &   1.7$+$08  \\
    6  &  18   &  9.4$-$02   &  8.0$+$09  &  8.1$+$09   &   8.0$+$09  &  8.0$+$09  & 7.8$+$09	&   7.4$+$09  &   7.8$+$09  \\
    6  &  19   &  1.8$-$02   &  7.8$+$08  &  7.9$+$08   &   7.8$+$08  &  7.8$+$08  & 7.3$+$08	&   7.4$+$08  &   7.8$+$08  \\
    6  &  31   &  2.1$-$01   &  1.0$+$10  &  1.0$+$10   &   1.0$+$10  &  1.0$+$10  & 1.0$+$10	&   1.0$+$10  &   9.6$+$09  \\
    6  &  36   &  3.3$-$01   &  2.3$+$10  &  2.3$+$10   &   2.2$+$10  &  2.2$+$10  & 2.3$+$10	&   2.1$+$10  &   2.1$+$10  \\
    6  &  37   &  6.4$-$04   &  1.4$+$08  &  1.4$+$08   &   1.5$+$08  &  1.4$+$08  &	     	&   3.3$+$08  &   1.9$+$08  \\
    7  &  13   &  8.7$-$03   &  3.5$+$08  &  3.5$+$08   &   3.5$+$08  &  3.5$+$08  & 3.7$+$08	&   1.9$+$08  &   3.7$+$08  \\
    7  &  14   &  6.2$-$02   &  1.7$+$09  &  1.7$+$09   &   1.7$+$09  &  1.7$+$09  & 1.7$+$09	&   1.8$+$09  &   1.9$+$09  \\
    7  &  19   &  1.1$-$01   &  6.8$+$09  &  6.8$+$09   &   6.8$+$09  &  6.7$+$09  & 6.4$+$09	&   6.6$+$09  &   6.4$+$09  \\
    7  &  32   &  2.0$-$01   &  1.2$+$10  &  1.2$+$10   &   1.2$+$10  &  1.2$+$10  & 1.1$+$10	&   1.1$+$10  &   1.1$+$10  \\
    7  &  35   &  3.1$-$01   &  2.4$+$10  &  2.4$+$10   &   2.4$+$10  &  2.4$+$10  & 2.4$+$10	&   2.3$+$10  &   2.3$+$10  \\
    7  &  36   &  1.4$-$02   &  1.4$+$09  &  1.4$+$09   &   1.4$+$09  &  1.4$+$09  & 1.4$+$09	&   1.5$+$08  &   1.3$+$09  \\
    8  &  18   &  1.2$-$02   &  3.3$+$08  &  3.2$+$08   &   3.2$+$08  &  3.2$+$08  & 3.5$+$08	&   7.0$+$08  &   3.9$+$08  \\
    8  &  33   &  7.6$-$01   &  2.8$+$10  &  2.8$+$10   &   2.8$+$10  &  2.7$+$10  & 2.7$+$10	&   2.5$+$10  &   2.5$+$10  \\
    8  &  34   &  2.5$-$01   &  1.8$+$10  &  1.8$+$10   &   1.8$+$10  &  1.8$+$10  & 1.9$+$10	&   2.7$+$10  &   1.6$+$10  \\
    9  &  18   &  7.2$-$02   &  1.5$+$09  &  1.6$+$09   &   1.5$+$09  &  1.5$+$09  & 1.6$+$09	&   1.1$+$09  &   1.1$+$09  \\
    9  &  33   &  3.1$-$02   &  1.1$+$09  &  1.0$+$09   &   1.0$+$09  &  1.1$+$09  & 1.1$+$09	&   2.1$+$09  &   1.1$+$09  \\
   10  &  18   &  1.1$-$02   &  4.6$+$08  &  4.7$+$08   &   4.7$+$08  &  4.7$+$08  & 4.9$+$08	&   5.0$+$08  &   4.9$+$08  \\
   10  &  19   &  7.0$-$02   &  1.4$+$09  &  1.4$+$09   &   1.4$+$09  &  1.4$+$09  & 1.5$+$09	&   1.2$+$09  &   1.6$+$09  \\
   10  &  33   &  1.4$-$01   &  8.9$+$09  &  8.7$+$09   &   8.7$+$09  &  8.7$+$09  & 8.1$+$09	&   9.8$+$09  &   8.5$+$09  \\
   10  &  34   &  2.7$-$02   &  3.6$+$09  &  3.5$+$09   &   3.5$+$09  &  3.5$+$09  & 3.2$+$09	&   3.1$+$09  &   3.6$+$09  \\
   10  &  37   &  4.9$-$02   &  7.1$+$09  &  7.2$+$09   &   7.1$+$09  &  7.0$+$09  & 7.1$+$09	&   7.1$+$09  &   5.5$+$09  \\
   10  &  40   &  9.0$-$01   &  4.6$+$10  &  4.6$+$10   &   4.5$+$10  &  4.5$+$10  & 4.4$+$10	&   4.6$+$10  &   4.3$+$10  \\
\hline            								                	 
\end{tabular}  

\begin{tabular}{rrrrrrrrrr} \hline
I & J   &      f (GRASP2)  & GRASP1     & GRASP2      & FAC1        & FAC2        &  MCHF	  &  CIV3     & NIST	    \\  
 \hline
   11  &  31   &  5.3$-$02   &  9.4$+$08  &  9.3$+$08   &   9.3$+$08  &  9.4$+$08  &	     	&   1.0$+$10  &   1.1$+$09  \\
   11  &  36   &  5.0$-$01   &  1.5$+$10  &  1.5$+$10   &   1.5$+$10  &  1.4$+$10  &	     	&   1.5$+$10  &   1.6$+$10  \\
   11  &  37   &  2.0$-$01   &  2.1$+$10  &  2.0$+$10   &   2.0$+$10  &  2.0$+$10  &	     	&   1.9$+$10  &   2.1$+$10  \\
   11  &  40   &  1.8$-$02   &  6.0$+$08  &  6.0$+$08   &   6.0$+$08  &  5.8$+$08  &	     	&   8.5$+$08  &   5.4$+$08  \\
   12  &  32   &  5.9$-$02   &  1.2$+$09  &  1.2$+$09   &   1.2$+$09  &  1.2$+$09  &	     	&   1.3$+$09  &   1.5$+$09  \\
   12  &  33   &  3.3$-$04   &  4.0$+$07  &  2.0$+$07   &   2.4$+$07  &  3.5$+$07  &	     	&   2.0$+$07  &   8.2$+$07  \\
   12  &  35   &  4.6$-$01   &  1.5$+$10  &  1.5$+$10   &   1.5$+$10  &  1.5$+$10  &	     	&   1.5$+$10  &   1.8$+$10  \\
   12  &  40   &  2.5$-$01   &  1.3$+$10  &  1.3$+$10   &   1.3$+$10  &  1.3$+$10  &	     	&   1.2$+$10  &   1.3$+$10  \\
\hline            								                	 
\end{tabular}  
			      							   					       
\begin{flushleft}													       
{\small
GRASP1: present calculations from the {\sc grasp} code with 530 levels \\ 
GRASP2: present calculations from the {\sc grasp} code with 1387 levels\\ 
FAC1: present calculations from the {\sc fac} code with 1391 levels  \\  
FAC2: present calculations from the {\sc fac} code with 12,139 levels \\ 
MCHF: Froese$-$Fischer {\em et al} \cite{cff1} \\ 
CIV3: Singh {\em et al} \cite{mm1} \\ 
NIST: http://www.nist.gov/pml/data/asd.cfm \\								       
											     
}															       
\end{flushleft} 

\clearpage
\begin{flushleft}
Table 6. Comparison of radiative rates (A-values, s$^{-1}$) for transitions from 3s$^2$3p $^2$P$^o_{1/2,3/2}$\\ levels to higher excited levels of Ti X -- see Table 3 for level indices. $a{\pm}b \equiv a{\times}$10$^{{\pm}b}$. 
\end{flushleft}
\begin{tabular}{rrrrrrrrrr} \hline
 & & & & & & & & &  \\
I & J   &        f (GRASP2) & GRASP1     & GRASP2     & FAC1      & FAC2        \\
 \\ \hline
 1  &	47  &  1.1538$-$05  &  1.8398$+$06  &  1.9731$+$06  &  1.920$+$06  &  1.682$+$06  \\
 1  &	48  &  1.9776$-$05  &  6.7318$+$06  &  6.8196$+$06  &  6.619$+$06  &  5.720$+$06  \\
 1  &	51  &  2.4366$-$05  &  7.6795$+$06  &  8.5251$+$06  &  8.507$+$06  &  7.645$+$06  \\
 1  &	56  &  3.5023$-$04  &  6.4496$+$07  &  6.7085$+$07  &  6.606$+$07  &  6.255$+$07  \\
 1  &	60  &  5.3416$-$03  &  2.2535$+$10  &  2.2269$+$09  &  9.586$+$08  &  1.375$+$09  \\
 1  &	61  &  3.9163$-$04  &  4.2340$+$09  &  8.1951$+$07  &  7.066$+$07  &  8.034$+$07  \\
 1  &	62  &  4.7657$-$02  &  8.0402$+$07  &  1.9990$+$10  &  1.720$+$10  &  1.911$+$10  \\
 1  &	63  &  2.1532$-$04  &  3.8698$+$09  &  4.5669$+$07  &  4.216$+$07  &  4.440$+$07  \\
 1  &	64  &  1.9925$-$02  &  4.1529$+$07  &  8.4790$+$09  &  1.284$+$10  &  1.058$+$10  \\
 1  &	76  &  3.7376$-$04  &  9.4299$+$07  &  9.0881$+$07  &  7.855$+$07  &  7.495$+$07  \\
 1  &	82  &  1.3734$-$03  &  4.4977$+$08  &  3.5858$+$08  &  3.479$+$08  &  3.367$+$08  \\
 1  &	85  &  2.8067$-$05  &  4.2223$+$07  &  1.6713$+$07  &  1.633$+$07  &  7.181$+$06  \\
 1  &	86  &  5.2815$-$04  &  5.5242$+$08  &  3.1774$+$08  &  2.992$+$08  &  3.093$+$08  \\
 1  &	87  &  2.7039$-$04  &  1.4249$+$08  &  8.1434$+$07  &  7.730$+$07  &  7.976$+$07  \\
 1  &	92  &  1.7055$-$01  &  5.4164$+$10  &  5.5316$+$10  &  5.679$+$10  &  5.669$+$10  \\
 2  &	47  &  4.3171$-$05  &  1.4147$+$07  &  1.4459$+$07  &  1.437$+$07  &  1.251$+$07  \\
 2  &	48  &  3.8709$-$06  &  2.6820$+$06  &  2.6145$+$06  &  2.588$+$06  &  2.200$+$06  \\
 2  &	49  &  4.8757$-$06  &  1.4770$+$06  &  1.6626$+$06  &  1.626$+$06  &  1.446$+$06  \\
 2  &	51  &  4.4079$-$06  &  2.8412$+$06  &  3.0212$+$06  &  2.898$+$06  &  2.512$+$06  \\
 2  &	55  &  2.6298$-$04  &  6.5868$+$07  &  6.5618$+$07  &  6.302$+$07  &  5.983$+$07  \\
 2  &	56  &  3.6091$-$05  &  1.4046$+$07  &  1.3555$+$07  &  1.282$+$07  &  1.243$+$07  \\
 2  &	57  &  5.6168$-$05  &  1.0619$+$07  &  1.4244$+$07  &  1.379$+$07  &  1.302$+$07  \\
 2  &	59  &  7.7714$-$06  &  6.0428$+$06  &  5.9371$+$06  &  5.442$+$06  &  5.738$+$06  \\
 2  &	60  &  3.1447$-$03  &  4.2378$+$10  &  2.5727$+$09  &  8.171$+$08  &  1.354$+$09  \\
 2  &	61  &  7.2474$-$05  &  1.2664$+$10  &  2.9762$+$07  &  2.416$+$07  &  2.427$+$07  \\
 2  &	62  &  5.2419$-$02  &  6.4851$+$07  &  4.3150$+$10  &  3.656$+$10  &  4.080$+$10  \\
 2  &	63  &  1.6063$-$04  &  7.4092$+$09  &  6.6869$+$07  &  5.666$+$07  &  6.485$+$07  \\
 2  &	64  &  2.0234$-$02  &  6.6933$+$07  &  1.6901$+$10  &  2.589$+$10  &  2.121$+$10  \\
 2  &	65  &  4.2244$-$04  &  9.9079$+$07  &  1.1854$+$08  &  1.063$+$08  &  1.192$+$08  \\
 2  &	75  &  4.0113$-$04  &  1.3505$+$08  &  1.2708$+$08  &  1.107$+$08  &  1.071$+$08  \\
 2  &	76  &  3.5826$-$05  &  1.7738$+$07  &  1.7119$+$07  &  1.455$+$07  &  1.392$+$07  \\
 2  &	81  &  1.0755$-$03  &  4.6509$+$08  &  3.6805$+$08  &  3.555$+$08  &  3.428$+$08  \\
 2  &	82  &  1.4880$-$04  &  9.4412$+$07  &  7.6393$+$07  &  7.366$+$07  &  7.162$+$07  \\
 2  &	86  &  1.2823$-$04  &  2.6064$+$08  &  1.5188$+$08  &  1.434$+$08  &  1.435$+$08  \\
 2  &	87  &  6.3564$-$04  &  6.4515$+$08  &  3.7688$+$08  &  3.525$+$08  &  3.615$+$08  \\
 2  &	92  &  1.7643$-$02  &  1.1071$+$10  &  1.1272$+$10  &  1.155$+$10  &  1.152$+$10  \\
 2  &	93  &  1.5642$-$01  &  6.5482$+$10  &  6.6669$+$10  &  6.835$+$10  &  6.818$+$10  \\
 \hline            								                	 
\end{tabular}   								   					       
			      							   					       
\begin{flushleft}													       
{\small
GRASP1: present calculations from the {\sc grasp} code with 530 levels \\ 
GRASP2: present calculations from the {\sc grasp} code with 1387 levels\\ 
FAC1: present calculations from the {\sc fac} code with 1391 levels  \\  
FAC2: present calculations from the {\sc fac} code with 12,139 levels \\ 										
											     
}															       
\end{flushleft} 

\clearpage
\begin{flushleft}
Table 7. Comparison of lifetimes ($\tau$, ps) for the lowest 40 levels of Ti X. $a{\pm}b \equiv a{\times}$10$^{{\pm}b}$.
\end{flushleft}
{\tiny
\begin{tabular}{rlllllllllll}  \hline
 & & & & & & & & & & &  \\
Index  & \multicolumn{2}{c}{Configuration/Level} &  GRASP1 & GRASP2     & FAC1       & FAC2         &  CIV3      & MCHF      & MBPT      & Experimental$^a$   & Experimental$^b$                                            \\
& & & & & & & & & & &   \\ \hline
& & & & & & & & & & &   \\
  1  &  3s$^2$3p       &  $^2$P$^o$$_{1/2}$  &           &	       &	    &  	           &	        &		        &	               &  							           \\
  2  &  3s$^2$3p       &  $^2$P$^o$$_{3/2}$  & 2.673$+$11 & 2.673$+$11 & 2.703$+$11 &  2.701$+$11  &	     &  	 &	     &  		    &									\\
  3  &  3s3p$^2$       &  $^4$P$$$$$_{1/2}$  & 1.859$+$05 & 1.832$+$05 & 1.823$+$05 &  1.791$+$05  &	     &1.680$+$05 & 2.12$+$05 &  		    &									\\
  4  &  3s3p$^2$       &  $^4$P$$$$$_{3/2}$  & 9.303$+$05 & 9.197$+$05 & 9.144$+$05 &  8.992$+$05  &	     &8.895$+$05 & 1.09$+$05 &  		    &									\\
  5  &  3s3p$^2$       &  $^4$P$$$$$_{5/2}$  & 2.984$+$05 & 2.954$+$05 & 2.939$+$05 &  2.891$+$05  &	     &2.840$+$05 & 3.75$+$05 &  		    &									\\
  6  &  3s3p$^2$       &  $^2$D$$$$$_{3/2}$  & 8.007$+$02 & 7.999$+$02 & 8.035$+$02 &  7.965$+$02  &9.399$+$02  &7.624$+$02 & 9.21$+$02 & [8.20$\pm$0.30]$+$02 &	[8.90$\pm$0.40,9.20$\pm$0.50,8.40$\pm$0.30,8.50$\pm$0.60]$+$02  \\ 
  7  &  3s3p$^2$       &  $^2$D$$$$$_{5/2}$  & 9.023$+$02 & 9.017$+$02 & 9.058$+$02 &  8.985$+$02  &1.025$+$03  &8.579$+$02 & 1.05$+$03 & [8.70$\pm$0.40]$+$02 &	[9.70$\pm$0.30,9.80$\pm$0.50,8.50$\pm$0.60,9.50$\pm$0.50]$+$02  \\ 
  8  &  3s3p$^2$       &  $^2$S$$$$$_{1/2}$  & 9.487$+$01 & 9.555$+$01 & 9.647$+$01 &  9.676$+$01  &1.138$+$02  &1.012$+$02 & 1.08$+$02 & [9.90$\pm$0.60]$+$01 &	[1.15$\pm$0.80,9.40$\pm$0.50,1.20$\pm$0.50,1.09$\pm$0.10]$+$01  \\ 
  9  &  3s3p$^2$       &  $^2$P$$$$$_{1/2}$  & 3.711$+$01 & 3.740$+$01 & 3.768$+$01 &  3.766$+$01  &3.785$+$01  &3.809$+$01 & 4.22$+$01 & [6.00$\pm$0.15]$+$01 &	[6.00$\pm$0.40,6.60$\pm$0.50,6.70$\pm$0.30,4.30$\pm$0.50]$+$01  \\ 
 10  &  3s3p$^2$       &  $^2$P$$$$$_{3/2}$  & 3.537$+$01 & 3.565$+$01 & 3.591$+$01 &  3.593$+$01  &3.748$+$01  &3.652$+$01 & 4.02$+$01 & [5.00$\pm$0.15]$+$01 &	[5.60$\pm$0.80,4.60$\pm$0.50,5.00$\pm$0.30,3.40$\pm$0.50]$+$01  \\ 
 11  &  3s$^2$3d       &  $^2$D$$$$$_{3/2}$  & 3.072$+$01 & 3.079$+$01 & 3.117$+$01 &  3.126$+$01  &3.135$+$01  &     $ $   & 3.52$+$01 & [3.80$\pm$0.60]$+$01 &	[5.00$\pm$0.20,4.90$\pm$0.30,5.20$\pm$0.20,3.70$\pm$0.50]$+$01  \\ 
 12  &  3s$^2$3d       &  $^2$D$$$$$_{5/2}$  & 3.235$+$01 & 3.243$+$01 & 3.284$+$01 &  3.293$+$01  &3.311$+$01  &     $ $   & 3.72$+$01 & [3.60$\pm$0.50]$+$01 &	[5.90$\pm$0.20,5.80$\pm$0.30,5.90$\pm$0.20,4.40$\pm$0.60]$+$01  \\ 
 13  &  3p$^3$         &  $^2$D$^o$$_{3/2}$  & 4.684$+$02 & 4.662$+$02 & 4.672$+$02 &  4.661$+$02  &4.492$+$02  &4.349$+$02 &     $ $   & [4.70$\pm$0.50]$+$02 &									\\ 
 14  &  3p$^3$         &  $^2$D$^o$$_{5/2}$  & 4.693$+$02 & 4.672$+$02 & 4.684$+$02 &  4.673$+$02  &4.486$+$02  &4.380$+$02 &     $ $   & [4.90$\pm$0.40]$+$02 &									\\ 
 15  &  3p$^3$         &  $^4$S$^o$$_{3/2}$  & 4.110$+$01 & 4.098$+$01 & 4.117$+$01 &  4.124$+$01  &4.283$+$01  &4.146$+$01 &     $ $   &  		    &									\\
 16  &  3s3p($^3$P)3d  &  $^4$F$^o$$_{3/2}$  & 1.285$+$04 & 1.286$+$04 & 1.338$+$04 &  1.285$+$04  &     $ $    &8.226$+$03 & 1.77$+$04 &  		    &	[1.60$\pm$0.15]$+04$$^c$					\\
 17  &  3s3p($^3$P)3d  &  $^4$F$^o$$_{5/2}$  & 1.626$+$04 & 1.608$+$04 & 1.613$+$04 &  1.612$+$04  &     $ $    &1.740$+$04 & 1.88$+$04 &  		    &	[1.30$\pm$0.15]$+04$$^c$					\\
 18  &  3p$^3$         &  $^2$P$^o$$_{1/2}$  & 9.652$+$01 & 9.575$+$01 & 9.634$+$01 &  9.683$+$01  &1.030$+$02  &9.750$+$01 &     $ $   & [1.10$\pm$0.10]$+$02 &									\\ 
 19  &  3p$^3$         &  $^2$P$^o$$_{3/2}$  & 9.818$+$01 & 9.746$+$01 & 9.803$+$01 &  9.852$+$01  &1.057$+$02  &1.039$+$02 &     $ $   &  		    &									\\
 20  &  3s3p($^3$P)3d  &  $^4$F$^o$$_{7/2}$  & 1.874$+$04 & 1.851$+$04 & 1.856$+$04 &  1.861$+$04  &     $ $    &2.089$+$04 & 2.20$+$04 &  		    &	[1.85$\pm$0.20]$+04$$^c$					\\
 21  &  3s3p($^3$P)3d  &  $^4$F$^o$$_{9/2}$  & 7.181$+$10 & 7.174$+$10 & 7.285$+$10 &  7.319$+$10  &     $ $    &     $ $   &     $ $   &  		    &									\\
 22  &  3s3p($^3$P)3d  &  $^4$P$^o$$_{5/2}$  & 4.965$+$01 & 4.982$+$01 & 5.038$+$01 &  5.017$+$01  &5.718$+$01  &5.004$+$01 & 5.45$+$01 &  		    &									\\
 23  &  3s3p($^3$P)3d  &  $^4$P$^o$$_{3/2}$  & 4.383$+$01 & 4.401$+$01 & 4.448$+$01 &  4.407$+$01  &5.528$+$01  &4.393$+$01 & 4.76$+$01 &  		    &									\\
 24  &  3s3p($^3$P)3d  &  $^4$D$^o$$_{1/2}$  & 3.868$+$01 & 3.890$+$01 & 3.927$+$01 &  4.118$+$01  &3.132$+$01  &3.840$+$01 & 4.51$+$01 &  		    &									\\
 25  &  3s3p($^3$P)3d  &  $^4$P$^o$$_{1/2}$  & 3.974$+$01 & 3.970$+$01 & 4.027$+$01 &  3.857$+$01  &5.426$+$01  &4.127$+$01 & 4.18$+$01 &  		    &									\\
 26  &  3s3p($^3$P)3d  &  $^4$D$^o$$_{3/2}$  & 3.622$+$01 & 3.625$+$01 & 3.672$+$01 &  3.712$+$01  &3.153$+$01  &3.707$+$01 & 4.05$+$01 &  		    &									\\
 27  &  3s3p($^3$P)3d  &  $^4$D$^o$$_{5/2}$  & 3.398$+$01 & 3.404$+$01 & 3.445$+$01 &  3.464$+$01  &3.175$+$01  &3.449$+$01 & 3.80$+$01 &  		    &									\\
 28  &  3s3p($^3$P)3d  &  $^4$D$^o$$_{7/2}$  & 3.220$+$01 & 3.227$+$01 & 3.265$+$01 &  3.272$+$01  &3.226$+$01  &3.256$+$01 & 3.59$+$01 &  		    &									\\
 29  &  3s3p($^3$P)3d  &  $^2$D$^o$$_{3/2}$  & 3.835$+$01 & 3.833$+$01 & 3.868$+$01 &  3.886$+$01  &4.098$+$01  &3.972$+$01 &     $ $   & [5.00$\pm$0.30]$+$01 &									\\ 
 30  &  3s3p($^3$P)3d  &  $^2$D$^o$$_{5/2}$  & 3.859$+$01 & 3.857$+$01 & 3.893$+$01 &  3.912$+$01  &4.109$+$01  &4.003$+$01 &     $ $   & [5.00$\pm$0.30]$+$01 &									\\ 
 31  &  3s3p($^3$P)3d  &  $^2$F$^o$$_{5/2}$  & 7.824$+$01 & 7.838$+$01 & 7.933$+$01 &  8.036$+$01  &8.537$+$01  &8.911$+$01 &     $ $   & [8.20$\pm$0.60]$+$01 &									\\ 
 32  &  3s3p($^3$P)3d  &  $^2$F$^o$$_{7/2}$  & 7.567$+$01 & 7.581$+$01 & 7.672$+$01 &  7.774$+$01  &8.157$+$01  &8.682$+$01 &     $ $   & [8.80$\pm$0.80]$+$01 &									\\ 
 33  &  3s3p($^3$P)3d  &  $^2$P$^o$$_{3/2}$  & 2.637$+$01 & 2.637$+$01 & 2.670$+$01 &  2.671$+$01  &2.695$+$01  &2.765$+$01 &     $ $   & [3.40$\pm$1.00]$+$01 &									\\ 
 34  &  3s3p($^3$P)3d  &  $^2$P$^o$$_{1/2}$  & 2.766$+$01 & 2.769$+$01 & 2.804$+$01 &  2.806$+$01  &2.706$+$01  &2.902$+$01 &     $ $   & [4.70$\pm$0.40]$+$01 &									\\ 
 35  &  3s3p($^1$P)3d  &  $^2$F$^o$$_{7/2}$  & 2.550$+$01 & 2.556$+$01 & 2.577$+$01 &  2.580$+$01  &2.638$+$01  &4.200$+$01 & 3.03$+$01 & [3.70$\pm$0.30]$+$01 &									\\ 
 36  &  3s3p($^1$P)3d  &  $^2$F$^o$$_{5/2}$  & 2.525$+$01 & 2.531$+$01 & 2.552$+$01 &  2.556$+$01  &2.616$+$01  &4.210$+$01 & 3.00$+$01 &  		    &									\\
 37  &  3s3p($^1$P)3d  &  $^2$P$^o$$_{1/2}$  & 2.154$+$01 & 2.174$+$01 & 2.196$+$01 &  2.226$+$01  &2.273$+$01  &4.045$+$01 & 3.57$+$01 & [4.90$\pm$1.30]$+$01 &									\\ 
 38  &  3s3p($^1$P)3d  &  $^2$P$^o$$_{3/2}$  & 2.212$+$01 & 2.234$+$01 & 2.253$+$01 &  2.286$+$01  &2.311$+$01  &4.267$+$01 & 5.56$+$01 & [4.90$\pm$1.30]$+$01 &									\\ 
 39  &  3s3p($^1$P)3d  &  $^2$D$^o$$_{3/2}$  & 1.683$+$01 & 1.681$+$01 & 1.700$+$01 &  1.720$+$01  &1.670$+$01  &2.315$+$01 & 2.11$+$01 & [2.70$\pm$0.70]$+$01 &									\\ 
 40  &  3s3p($^1$P)3d  &  $^2$D$^o$$_{5/2}$  & 1.681$+$01 & 1.680$+$01 & 1.698$+$01 &  1.716$+$01  &1.699$+$01  &2.279$+$01 & 1.94$+$01 & [2.70$\pm$0.70]$+$01 &									\\ 
& & & & & & & & & & &  \\ \hline            								                	 
\end{tabular}   								   					       
			      							   					       
\begin{flushleft}													       
{\small
GRASP1: present calculations from the {\sc grasp} code with 530 levels \\
GRASP2: present calculations from the {\sc grasp} code with 1387 levels \\
FAC1: present calculations from the {\sc fac} code with 1391 levels \\  
FAC2: calculations with 12,139 levels \\ 													       
CIV3: Singh {\em et al} \cite{mm2}  \\
MCHF: Froese$-$Fischer {\em et al} \cite{cff1} \\
MBPT: Safronova {\em et al}  \cite{uis}  \\
a: Pinnington {\em et al}  \cite{ehp1} \\
b: Pinnington {\em et al}  \cite{ehp2}, the first entry is for Free M-E Fit, the second for Constrained M-E Fit, the 3rd for VNET, and the 4th for ANDC \\ 
c: Tr{\" a}bert {\em et al} \cite{et1} \\
}															       
\end{flushleft} 
}													       

\clearpage
\begin{flushleft}
{\bf Table 8.}  Comparison of lifetimes ($\tau$, ps) for the 3s3p4$\ell$ levels of Ti X. 
\end{flushleft}
\begin{tabular}{rllrrrllll} \hline
 & & & &   & & & & \\
Index  & \multicolumn{2}{c}{Configuration/Level} & GRASP & CIV3	& Index & \multicolumn{2}{c}{Configuration/Level}   & GRASP   & CIV3     \\
& & & &   & & & &  \\ \hline
& & & &   & & & &  \\
   1  &  3s3p($^3$P)4s  & $^4$P$^o_{ 1/2 }$   &  13.86  &  	    &   37  &  3s3p($^3$P)4d  & $^4$P$^o_{ 5/2 }$   &  16.04  &   17.54  \\
   2  &  3s3p($^3$P)4s  & $^4$P$^o_{ 3/2 }$   &  13.85  &  	    &   38  &  3s3p($^3$P)4d  & $^4$P$^o_{ 3/2 }$   &  16.17  &   17.54  \\
   3  &  3s3p($^3$P)4s  & $^4$P$^o_{ 5/2 }$   &  13.92  &  	    &   39  &  3s3p($^3$P)4d  & $^4$P$^o_{ 1/2 }$   &  16.36  &   17.15  \\
   4  &  3s3p($^3$P)4s  & $^2$P$^o_{ 1/2 }$   &  12.07  &  	    &   40  &  3s3p($^3$P)4d  & $^2$F$^o_{ 7/2 }$   &  12.36  &   13.19  \\
   5  &  3s3p($^3$P)4s  & $^2$P$^o_{ 3/2 }$   &  11.58  &  	    &   41  &  3s3p($^3$P)4d  & $^2$P$^o_{ 3/2 }$   &  16.49  &   18.69  \\
   6  &  3s3p($^1$P)4s  & $^2$P$^o_{ 1/2 }$   &  10.38  &    7.08   &   42  &  3s3p($^3$P)4d  & $^2$P$^o_{ 1/2 }$   &  16.03  &   19.08  \\
   7  &  3s3p($^1$P)4s  & $^2$P$^o_{ 3/2 }$   &  10.79  &   12.87   &   43  &  3s3p($^1$P)4d  & $^2$F$^o_{ 5/2 }$   &  12.70  &          \\
   8  &  3s3p($^3$P)4p  & $^4$D$  _{ 1/2 }$   &  21.63  &  	    &   44  &  3s3p($^1$P)4d  & $^2$F$^o_{ 7/2 }$   &  13.79  &          \\
   9  &  3s3p($^3$P)4p  & $^4$D$  _{ 3/2 }$   &  21.98  &  	    &   45  &  3s3p($^1$P)4d  & $^2$D$^o_{ 3/2 }$   &	9.90  &          \\
  10  &  3s3p($^3$P)4p  & $^2$P$  _{ 1/2 }$   &  19.36  &  	    &   46  &  3s3p($^1$P)4d  & $^2$D$^o_{ 5/2 }$   &	9.91  &          \\
  11  &  3s3p($^3$P)4p  & $^2$P$  _{ 3/2 }$   &  20.08  &   14.75   &   47  &  3s3p($^1$P)4d  & $^2$P$^o_{ 1/2 }$   &  17.05  &          \\
  12  &  3s3p($^3$P)4p  & $^4$D$  _{ 5/2 }$   &  37.64  &  	    &   48  &  3s3p($^1$P)4d  & $^2$P$^o_{ 3/2 }$   &  17.14  &          \\
  13  &  3s3p($^3$P)4p  & $^4$D$  _{ 7/2 }$   &  37.46  &   42.37   &   49  &  3s3p($^3$P)4f  & $^4$G$  _{ 5/2 }$   & 4.56    &          \\
  14  &  3s3p($^3$P)4p  & $^4$P$  _{ 1/2 }$   &  52.27  &   60.94   &   50  &  3s3p($^3$P)4f  & $^4$F$  _{ 7/2 }$   & 4.62    &          \\
  15  &  3s3p($^3$P)4p  & $^4$P$  _{ 3/2 }$   &  48.42  &   62.50   &   51  &  3s3p($^3$P)4f  & $^4$F$  _{ 3/2 }$   & 4.79    &          \\
  16  &  3s3p($^3$P)4p  & $^4$P$  _{ 5/2 }$   &  49.85  &   62.11   &   52  &  3s3p($^3$P)4f  & $^4$F$  _{ 9/2 }$   & 4.65    &          \\
  17  &  3s3p($^3$P)4p  & $^4$S$  _{ 3/2 }$   &  40.24  &   42.55   &   53  &  3s3p($^3$P)4f  & $^4$F$  _{ 5/2 }$   & 5.81    &          \\
  18  &  3s3p($^3$P)4p  & $^2$D$  _{ 3/2 }$   &  13.90  &   49.95   &   54  &  3s3p($^3$P)4f  & $^4$G$  _{ 7/2 }$   & 4.57    &          \\
  19  &  3s3p($^3$P)4p  & $^2$D$  _{ 5/2 }$   &  14.04  &   47.62   &   55  &  3s3p($^3$P)4f  & $^4$G$  _{ 9/2 }$   & 4.51    &          \\
  20  &  3s3p($^3$P)4p  & $^2$S$  _{ 1/2 }$   &  13.91  &  	    &   56  &  3s3p($^3$P)4f  & $^4$G$  _{ 11/2}$   & 4.40    &          \\
  21  &  3s3p($^1$P)4p  & $^2$P$  _{ 1/2 }$   &  16.68  &    7.08   &   57  &  3s3p($^3$P)4f  & $^2$F$  _{ 5/2 }$   & 4.69    &  5.48    \\
  22  &  3s3p($^1$P)4p  & $^2$D$  _{ 3/2 }$   &  20.29  &  	    &   58  &  3s3p($^3$P)4f  & $^2$F$  _{ 7/2 }$   & 4.68    &  5.48    \\
  23  &  3s3p($^1$P)4p  & $^2$D$  _{ 5/2 }$   &  22.41  &  	    &   59  &  3s3p($^3$P)4f  & $^4$D$  _{ 7/2 }$   & 4.64    &          \\
  24  &  3s3p($^1$P)4p  & $^2$P$  _{ 3/2 }$   &  18.42  &   12.87   &   60  &  3s3p($^3$P)4f  & $^4$D$  _{ 5/2 }$   & 4.63    &          \\
  25  &  3s3p($^1$P)4p  & $^2$S$  _{ 1/2 }$   &  17.40  &  	    &   61  &  3s3p($^3$P)4f  & $^4$D$  _{ 3/2 }$   & 4.62    &          \\
  26  &  3s3p($^3$P)4d  & $^4$D$^o_{ 1/2 }$   &   8.71  &  	    &   62  &  3s3p($^3$P)4f  & $^4$D$  _{ 1/2 }$   & 4.62    &          \\
  27  &  3s3p($^3$P)4d  & $^4$D$^o_{ 3/2 }$   &   8.99  &  	    &   63  &  3s3p($^3$P)4f  & $^2$G$  _{ 7/2 }$   & 5.32    &  6.44    \\
  28  &  3s3p($^3$P)4d  & $^4$D$^o_{ 5/2 }$   &   9.68  &  	    &   64  &  3s3p($^3$P)4f  & $^2$G$  _{ 9/2 }$   & 5.36    &  4.70    \\
  29  &  3s3p($^3$P)4d  & $^2$D$^o_{ 3/2 }$   &  24.55  &   68.49   &   65  &  3s3p($^3$P)4f  & $^2$D$  _{ 5/2 }$   & 5.39    &          \\
  30  &  3s3p($^3$P)4d  & $^4$D$^o_{ 7/2 }$   &   9.16  &  	    &   66  &  3s3p($^3$P)4f  & $^2$D$  _{ 3/2 }$   & 5.43    &          \\
  31  &  3s3p($^3$P)4d  & $^2$D$^o_{ 5/2 }$   &  21.54  &   69.21   &   67  &  3s3p($^1$P)4f  & $^2$F$  _{ 5/2 }$   & 4.42    &          \\
  32  &  3s3p($^3$P)4d  & $^4$F$^o_{ 3/2 }$   & 112.90  &  208.33   &   68  &  3s3p($^1$P)4f  & $^2$F$  _{ 7/2 }$   & 4.43    &          \\
  33  &  3s3p($^3$P)4d  & $^4$F$^o_{ 5/2 }$   &  57.23  &  229.36   &   69  &  3s3p($^1$P)4f  & $^2$D$  _{ 3/2 }$   & 4.49    & 11.05    \\
  34  &  3s3p($^3$P)4d  & $^4$F$^o_{ 7/2 }$   &  73.05  &  232.56   &   70  &  3s3p($^1$P)4f  & $^2$D$  _{ 5/2 }$   & 4.45    &  7.25    \\
  35  &  3s3p($^3$P)4d  & $^4$F$^o_{ 9/2 }$   & 251.40  &  233.73   &   71  &  3s3p($^1$P)4f  & $^2$G$  _{ 9/2 }$   & 7.50    &          \\
  36  &  3s3p($^3$P)4d  & $^2$F$^o_{ 5/2 }$   &  13.82  &   14.62   &   72  &  3s3p($^1$P)4f  & $^2$G$  _{ 7/2 }$   & 7.59    &          \\
& & & &   & & & &   \\ \hline            					        					
\end{tabular}   							        					      
			      						        					      
\begin{flushleft}							        					      
{\small 								      
GRASP: present calculations from the {\sc grasp} code with 1387 levels\\     
CIV3: calculations of Singh {\em et al} \cite{mm2} from the CIV3 code  \\ 
															       
}															       
\end{flushleft} 
\end{document}